\documentclass[a4paper,fleqn,usenatbib,useAMS]{mnras}

\usepackage{graphicx}	
\usepackage{amsmath}	
\usepackage{amssymb}	
\usepackage{multicol}        
\usepackage{bm}		
\usepackage{pdflscape}	

\usepackage[T1]{fontenc}
\usepackage{ae,aecompl}

\title{Prompt Gamma Ray Burst emission from gradual magnetic dissipation}
\author[P. Beniamini, D. Giannios]{Paz Beniamini$^{1,2}$ and Dimitrios Giannios$^3$
	\\
	$^1$Institut d'Astrophysique de Paris UMR 7095 Universit\'{e}
	Pierre et Marie Curie-Paris 06; CNRS 98 bis bd Arago, 75014 Paris, France \\
	$^2$Department of Physics, The George Washington University, Washington, DC 20052, USA \\
	$^3$Department of Physics and Astronomy, Purdue University, 525 Northwestern Avenue, West Lafayette, IN 47907, USA}

\begin{document}
\label{firstpage}
\pagerange{\pageref{firstpage}--\pageref{lastpage}}
\maketitle

\begin{abstract}
We considered a model for the prompt phase of Gamma-Ray Burst (GRB) emission arising from a magnetized jet undergoing gradual energy dissipation due to magnetic reconnection. The dissipated magnetic energy is translated to bulk kinetic energy and to acceleration of particles. The energy in these particles is released via synchrotron radiation as they gyrate around the strong magnetic fields in the jet. At small radii, the optical depth is large, and the radiation is reprocessed through Comptonization into a narrow, strongly peaked, component. At larger distances the optical depth becomes small and radiation escapes the jet with a non-thermal distribution.
The obtained spectra typically peak around $\approx 300$keV (as observed) and with spectral indices below and above the peak that are, for a broad range of the model parameters, close to the observed values. The small radius of dissipation causes the emission to become self absorbed at a few keV and can sufficiently suppress the optical and X-ray fluxes within the limits required by observations \citep{Beniamini2014}.
\end{abstract}

\begin{keywords}
 gamma-ray burst: general
\end{keywords}

\section{Introduction}
Relativistic Poynting-flux dominated jets have been suggested to occur in various astrophysical outflows, including active galactic nuclei (AGN), pulsars, micro-quasars and Gamma-Ray bursts (GRBs). The alternative situation, that the jet is accelerated through thermal pressure gradients, has been ruled out for AGNs, based on the available thermal power at the jet's base \citep{Ghisellini2009}.
This suggests that magnetic jets dominate those environments and indeed may be universal, thus encouraging us to consider the possibility of magnetic jets operating in GRBs as well (see also \citealt{Leng2014}). Furthermore, magnetic jets in GRBs are supported based on modelling of the accretion disc \citep{Kawanaka2013}.

The emission process in prompt GRBs is highly debated.
The overall non-thermal spectrum has led to the suggestion that the radiation is dominated by synchrotron emission from a power-law distribution of electrons \citep{Katz1994,Rees1994,Sari1996,Sari1998,Kumar2008,Daigne2011,Beniamini2013}.
A major concern for this model regards the low energy spectral slope.
The synchrotron fast cooling (which is the required cooling regime for obtaining large radiative efficiencies) 
photon index below the peak, $dN/d\nu\equiv N_{\nu}\propto \nu^{\alpha}$ is $\alpha=-1.5$, and is inconsistent with the typically observed slope of
$\alpha=-1$. Although it is relatively easy to achieve softer spectra $\alpha<-1.5$, it is extremely difficult to increase $\alpha$ in these models \footnote{Inverse Compton (IC) cooling in the Klein Nishina (KN) regime could help to increase somewhat the low energy synchrotron spectral slope.
At best, if only a small fraction of the electrons are accelerated, the slope could approach $\alpha \to -1$ \citep{Daigne2011}.} 
However, $90\%$ of all GRB have $\alpha>-1.5$ \citep{Preece2000, Ghirlanda2002, Kaneko2006, Nava2011} and
$\approx 40\%$ of GRBs  have $\alpha>-2/3$ \citep{Nava2011}, which is impossible for optically thin synchrotron, even in the
slow cooling regime. This problem is referred to as the synchrotron ``line of death'' \citep{Preece1998}.
Other problems with a synchrotron origin for the prompt phase involve the narrow peak energy distribution and the apparent sharp decline of the peak energy distribution above 1 MeV \citep{Band1993, Mallozzi1995,Schaefer2003, Beloborodov2013}, although this could also be due to a selection effect \citep{Shahmoradi2010, Beniamini2013}, and the narrow spectral width of the observed ``Band function" as compared with the synchrotron peak \citep{Baring2004,Burgess2011,Daigne2011,Beloborodov2013,Yu2016}.
The main alternative to synchrotron is photospheric emission. The photospheric emission component can be powerful. It is characterized by hard low-energy spectral slopes and, provided that there is energy dissipation close to the photosphere, a broader spectrum that resembles the observed one
\citep{Goodman1986,Thompson1994,Meszaros2000,Giannios2006,Beloborodov2010,Lazzati2010,Giannios2012}.
Photospheric emission models have their own share of problems, such as the need for a significant amount of dissipation and non-thermal acceleration deep below the photosphere \citep{Vurm2013} and various GRBs in which strong limits have been put on the thermal component \citep{Ryde2009,Ryde2010,Guiriec2011,Guiriec2013,Axelsson2012}.

In magnetic jets, the outflow may be composed of wound-up magnetic field lines in a ``striped wind" configuration \citep{Coroniti1990,Spruit2001}, in which their polarity is reversed over a typical distance $\lambda$.
Reconnection is then the most efficient process for dissipating the magnetic energy and transferring it to the emitting particles \citep{Spruit2001,DS2002}.
Many studies \citep{Giannios2008,Zhang2011,McKinney2012,Sironi2015,BG2016,Granot2016,Kagan2016,Begue2017} have therefore recently considered reconnection models as strong candidates for reproducing the prompt sub-MeV emission.
In MHD jets, the energy dissipation through reconnection occurs gradually over a wide range of radii \citep{Drenkhahn2002}.
At small radii, the jet is optically thick and photons produced via synchrotron emission are thermalized, leading to, to a zeroth approximation, a black body-like signature from the photosphere \citep{GianniosSpruit2005}.
Assuming that the dissipated energy heats smoothly the flow (e.g., a slow heating scenario following \cite{Ghisellini1999}), \cite{Giannios2006} showed that Comptonization close to the photosphere distorts the spectrum resulting in a high-energy power-law tail.
At larger radii, the radiation is dominated by synchrotron emission, and the typical frequencies and fluxes evolve as a function of the magnetization, dissipation rate, etc. The emitted radiation SED from both the $\tau \sim 1$ surface and the optically thin parts of the jet sensitively depends, however, on the assumptions on the injected particle spectrum at various scales in the jet.

Recent progress in understanding the non-thermal processes at the reconnection layer allow us to revisit this problem.
In particular, recent PIC simulations have demonstrated that magnetic reconnection in a high magnetization $\sigma$ (where $\sigma$ is the ratio of the Poynting flux to kinetic luminosities) plasma results in extended, power-law particle distribution. The particle power-law index depends sensitively on $\sigma$ and the particle spectra
become softer for lower $\sigma$ \citep[e.g.,][]{Cerutti2012,SironiSpitkovsky2014,Guo2014,Melzani2014,LloydRonning2016,SGP2016}.
These studies make specific predictions on the distribution of the accelerated particles expected at different scales in the jet.

Previous works have focused on one or just a few of these various ingredients.
In this work we combine the MHD dynamics, the details of the dissipation process as implied by new results from PIC simulations and a simple calculation for the emerging emission within a self-consistent model.
This model has a rather small number of free parameters: The jet luminosity, $L$, the ``wavelength" of the structured magnetic field in the jet over the outflow velocity from the reconnection sites, $\lambda/\epsilon$, the initial total energy per Baryon, $\eta$, and the fraction of accelerated electrons, $\xi$. The dynamical ranges of these parameters are expected to be quite small.
It is naturally expected to result in spectra peaking around $\approx 300$keV (as observed) and with spectral indices that closely match the observed values. In addition, the small radius of dissipation causes the spectrum to become self absorbed at a few keV and can sufficiently suppress the optical and X-ray fluxes below the upper limits required by observations \citep{Beniamini2014}.
\section{Model description}
\subsection{Dynamics}
\label{sec:dynamics}
We consider a Poynting flux dominated outflow that is composed of a ``striped wind" magnetic field configuration \citep{Coroniti1990,Spruit2001}, in which the typical ``wavelength" of the field is $\lambda$.
Such a configuration can be expected either if the central engine is a non-axisymmetric rotating millisecond magnetar or if it is an accreting BH. In the millisecond magnetar model, the scale of the stripe, $\lambda$, is related to the frequency of the rotator $\lambda \sim \pi c/\Omega \sim 10^8$ cm. For an accreting BH,
where the magnetic field that is advected inwards with the flow randomly switches sign (e.g. due to an instability in the disc, or accretion of different blobs of plasma with randomly oriented frozen-in field lines), $\lambda/c$ corresponds to the time-scale to accrete  a blob that forms close to the inner edge of the disc.
Recent simulations of magnetic jets powered by BH accretion disks, provide motivation that the jet may be highly variable on time-scales: $10-10^3r_g/c$ \citep[e.g.][]{Parfrey2015},  where $r_g=GM/c^2$. Therefore, the length-scale $\lambda=10^7-10^9$cm is typical for both GRB central engine models.

As the flow propagates, magnetic energy is dissipated via reconnection of oppositely oriented field lines.
The energy dissipation rate governs the dynamics and radiation and is determined by the inflow velocity of the plasma towards the reconnection layer, $v_{rec}\equiv\epsilon c$. Results from both analytical studies \citep{Lyubarsky2005} and PIC simulations \citep{Guo2015,Liu2015}, suggest typical values $\epsilon=0.1-0.25$.
For the purposes of the jet dynamics, $\lambda,\epsilon$ appear only through the specific combination $\lambda/\epsilon$ \citep{DS2002}.
We canonically choose $\lambda/\epsilon=10^8$cm (which for instance could be realized with $\lambda=50r_g(3M_{\odot})$, $\epsilon=0.2$).
The uncertainty in this parameter can also be estimated.
Given the ranges of expected values for $\lambda\ \& \ \epsilon$ noted above, we typically expect $3\times 10^7 \mbox{cm}\lesssim \lambda/\epsilon \lesssim 3\times 10^9$cm.

We consider a steady flow, characterized by a luminosity $L$.
The flow at the Alfven point is characterized by its magnetization parameter:
\begin{equation}
\sigma_0=\frac{L_{B,0}}{L_{kin,0}}=\frac{\beta_0 c(B_0 r_0)^2}{4\pi \Gamma_0 \dot{M} c^2}
\end{equation}
where $\beta=v/c$, $B_0$ and $\Gamma_0$ are the magnetic field strength and bulk Lorentz factor at the radius $r_0$ and $\dot{M}$ is the mass flux per sterad (all in the central engine frame).

The flow starts at $r_0$ with the Alfven speed $\Gamma_0=\sqrt{\sigma_0+1}\gg 1$ and accelerates as $\sigma$ decreases (so we can assume $\beta \approx 1$). The total luminosity (per steradian) at this radius is the magnetic plus kinetic components:
\begin{equation}
\label{eq:L}
L=(\sigma_0+1)\Gamma_0 \dot{M} c^2\approx\sigma_0 \Gamma_0 \dot{M} c^2.
\end{equation}
Assuming that a fraction of order unity of the dissipated energy goes towards accelerating the flow, we have that at any given radius $\sigma(r)\Gamma(r)=\sigma_0 \Gamma_0$. The acceleration is completed when $\sigma \lesssim 1$ and we get the terminal Lorentz factor (or equivalently the total energy per Baryon, $\eta$) $\eta \equiv \Gamma_{\infty}=\Gamma_0 \sigma_0=\sigma_0^{3/2}$. We denote the location where this happens by $r_s$ (saturation radius). The dependence of $\Gamma$ on radius for $r_0<r<r_s$ is derived
in \cite{Drenkhahn2002}, and found to be:
\begin{equation}
\label{eq:Gamma}
\Gamma=\Gamma_{\infty} \bigg(\frac{r}{r_s}\bigg)^{1/3}
\end{equation} 
where $r_s=\lambda \Gamma_{\infty}^2/(6 \epsilon)= 1.7\times 10^{13}\eta_3^2(\lambda/\epsilon)_8$cm.
Intuitively, $r_s$ can be understood as the radius by which particles that were initially furthest away from the reconnection sheets, finally reach the reconnection layer, where they are accelerated.

Following \cite{GianniosSpruit2005} we can now compute how the various flow parameters change with radius. The Poynting luminosity (per steradian) at any radius is given by:
\begin{equation}
\label{eq:LB}
L_B=c\frac{(rB)^2}{4\pi}=L\bigg(1-\frac{\Gamma}{\Gamma_{\infty}}\bigg)
\end{equation}
From this we obtain the energy dissipation rate, $d\dot{E}$, released between $r$ and $r+dr$:
\begin{equation}
d\dot{E}=-\frac{dL_B}{dr} dr=\frac{0.15L}{r_{12}^{2/3}}\eta_{2.5}^{-2/3}\bigg(\frac{\lambda}{\epsilon}\bigg)_8^{-1/3}dr_{12}
\end{equation}
where we use here and elsewhere the notation $q_x$ for $q=10^x$ in cgs units.
Therefore $\dot{E} = \int d\dot{E} \propto r^{1/3}$. We see that the energy dissipation is quite gradual and although it is dominated by the outer
regions of the flow ($r\to r_s$) it could still result in a significant contribution to the emission from the photospheric radius, $r_{ph}$ (to be defined below), depending on $(r_{ph}/r_s)$.

$d\dot{E}$ can be related to the electron number flux (assuming an electron-proton plasma):
\begin{equation}
d\dot{N}_e=\frac{d\dot{E}}{\xi^{-1}\Gamma \sigma m_p c^2}
\end{equation}
where $\xi$ is the fraction of these electrons accelerated in the reconnection sites and the denominator is the dissipated energy per accelerated particle in the flow.
As with $\dot{E}$, $\dot{N}_e(r)=\int d\dot{N}_e$ increases as $r^{1/3}$ (the rate of energy injection dictates the rate of particle injection).
In particular $\dot{N}_e(r_s)=\xi \dot{M}/m_p$ as is required by the fact that all of the available energy is dissipated by $r_s$.
PIC simulations suggest that $\xi$ is of order unity, i.e., a large fraction of the electrons undergo acceleration at the current sheet \citep{Sironi2015}. We assume here a range $0.03 \lesssim \xi \lesssim 1$ and use $\xi\approx 0.2$ as a canonical value.
PIC simulations find that the accelerated electrons initially form a power law energy spectrum (before particles can cool down due to radiation):
\begin{equation}
\frac{d\dot{N}_e}{d\gamma}=A\gamma^{-p} \quad ; \quad \gamma_i<\gamma<\gamma_f
\end{equation}
where $p$ depends sensitively on $\sigma$ \citep{SironiSpitkovsky2014,Guo2015,Kagan2015,Werner2016}.
\cite{SironiSpitkovsky2014} find $p=[4,3,2,1.5]$ for $\sigma=[1,3,10,50]$ correspondingly.
For numerical results we adopt: $p=4 \sigma^{-0.3}$ which results in a reasonable fit for the above figures, within the relevant range of $\sigma$.
The average energy per accelerated particle can now be written in terms of $\gamma_i,\gamma_f,p$.
This energy equals a fraction $\epsilon_e/2$ of the dissipated energy per particle (since about half of the dissipated energy goes directly towards bulk acceleration, see \citealt{DS2002}). We obtain (for $p\neq 1,2$):
\begin{equation}
\label{eq:gammas}
\frac{(1-p)}{(2-p)}\frac{\gamma_f^{2-p}-\gamma_i^{2-p}}{\gamma_f^{1-p}-\gamma_i^{1-p}}= \frac{\epsilon_e}{2 \xi}\sigma \frac{m_p}{m_e}.
\end{equation}
Reconnection simulations with electron-proton plasmas suggest that $\epsilon_e \approx 0.2$ \citep{Sironi2015}.  
Eq. \ref{eq:gammas} may be solved for $\gamma_i$ given $\gamma_f$, or vice versa.
For $p<2$ (corresponding to $\sigma>10$), $E_e$ is dominated by the highest energy particles and we can assume that $\gamma_i\approx 1$ (this is also in accordance with results of PIC simulations) and solve
Eq. \ref{eq:gammas} for $\gamma_f$. Since $\gamma_f\to \infty$ as $p\to 2$, this treatment thus becomes invalid once $\gamma>\gamma_{Max}$ (where $\gamma_{Max}$ is the maximal synchrotron energy obtained when the time for energy loss due to synchrotron equals the acceleration time, \citealt{deJager1996}). In this case, $\gamma_f=\gamma_{Max}$.
For $p>2$ (i.e. regions with $\sigma<10$), particles with $\gamma=\gamma_i$ dominate $E_e$ and Eq. \ref{eq:gammas} simplifies to $\gamma_i=\frac{p-2}{p-1}\frac{\epsilon_e}{2 \xi}\sigma \frac{m_p}{m_e}$.
In this regime we assume that $\gamma_f=\gamma_{Max}$ continues to hold, and solve Eq. \ref{eq:gammas} for $\gamma_i$ (in fact $\gamma_f$ quickly becomes irrelevant in this regime).

In order to determine the resulting emission from the flow, we turn to calculate the co-moving magnetic field and density.
Using Eqs. \ref{eq:Gamma}, \ref{eq:LB}, we obtain the co-moving magnetic field,
\begin{equation}
B'=\frac{B}{\Gamma}=\frac{4.1\times 10^6}{r_{12}^{4/3}}\frac{L_{52}^{1/2}}{\eta_3^{1/3}}\bigg(\frac{\lambda}{\epsilon}\bigg)_8^{1/3}\mbox{Gauss}.
\end{equation}
The co-moving density is obtained from the continuity equation $\dot{M}=r^2\Gamma \rho' c$, by plugging in Eqs. \ref{eq:Gamma}, \ref{eq:L},
\begin{equation}
\rho'=\frac{9.4\times10^{-10}}{r_{12}^{7/3}}\frac{L_{52}}{\eta_3^{4/3}}\bigg(\frac{\lambda}{\epsilon}\bigg)_8^{1/3}\mbox{g}/\mbox{cm}^3.
\end{equation}
We can relate the density to the optical depth as a function of radius by \citep{Abramowicz1991}
\begin{equation}
\tau=\int_r^\infty \Gamma (1-\beta)\kappa_{TS} \rho' dr
\end{equation}
where $\kappa_{TS}$ is the Thomson electron scattering opacity. The photospheric radius, is the radius for which $\tau(r_{ph})=1$,
\begin{equation}
\label{eq:rphot}
r_{ph}=4.6\times 10^{11} \bigg(\frac{\lambda}{\epsilon}\bigg)_8^{2/5} \frac{L_{52}^{3/5}}{\eta_3}\mbox{cm}.
\end{equation}
The photosphere could be pushed to larger radii if there is a significant amount of pair creation taking place at this radius. We return to this point in \S \ref{sec:rad}.
At the photospheric radius the Lorentz factor and magnetization are given by:
\begin{equation}
\Gamma(r_{ph})=\Gamma_{\infty} \bigg(\frac{r_{ph}}{r_s}\bigg)^{1/3}=300 L_{52}^{1/5}\bigg(\frac{\lambda}{\epsilon}\bigg)_8^{-1/5}
\end{equation}
\begin{equation}
\label{eq:sigmaph}
\sigma(r_{ph})=\eta/\Gamma(r_{ph})=3.2 \eta_3 L_{52}^{-1/5}\bigg(\frac{\lambda}{\epsilon}\bigg)_8^{1/5}.
\end{equation}
Thus, $\Gamma$ is almost fixed at the photosphere and of order $300$.
In addition, we see that the magnetization at the photosphere is mainly dependent on the baryon load.
Finally, notice that $r_{ph}<r_s$ requires
\begin{equation}
\eta>290 \bigg(\frac{\lambda}{\epsilon}\bigg)_8^{-1/5}L_{52}^{1/5}.
\end{equation}
which has a weak dependence on the model parameters.

\subsection{Radiation}
\label{sec:rad}
As energy is dissipated, it is transformed into radiation.
Energy is dissipated in both regions of high and low Thomson optical depth. Here we assume that sub-photospheric dissipation is reprocessed into a quasi-thermal, black body emission component while synchrotron dominates at the $\tau<1$ region.
In practice, neither the photospheric emission is a pure black body nor the optically thin emission pure synchrotron \citep[e.g.,][]{Giannios2006,GianniosSpruit2007,Beloborodov2011,Vurm2011,Beloborodov2013,Vurm2016}. The general tendency is that thermalization is achieved at large optical depths while the $\tau \ll 1$ region is dominated by synchrotron emission. However the transition does not take place suddenly at $\tau=1$. A full radiative transfer treatment of this problem is needed in order to properly analyse the intermediate regime. Such calculations have only been carried so far in the context of
photospheric models without energy dissipation close to the photosphere \citep{Ito2015,Lazzati2016}.
This requires a more involved calculation and is beyond the scope of the current work, in which we aim to highlight the qualitative features of the model at hand. 
We return briefly to discuss the validity of this simplifying assumption after describing the synchrotron frequencies and in \S \ref{sec:Comptonization}.

At small radii (below the photosphere), photons and matter are in thermodynamic equilibrium, both sharing some temperature $T$.
As the flow propagates, the co-moving temperature decreases: $T'\propto r^{-7/9}$ \citep{GianniosSpruit2005}. Since the thermal luminosity decreases as $L_{th}(r)\propto r^2\Gamma^2T'^4\propto r^{-4/9}$, only a fraction
$(r/r_{ph})^{4/9}$ of the dissipated energy at $r$ remains thermal at the photosphere. Plugging in the energy dissipation rate, we get
\begin{equation}
\begin{split}
\label{eq:Lph}
& L_{ph} \propto  \int_0^{r_{ph}}\frac{dr}{r^{2/3}} \bigg(\frac{r}{r_{ph}}\bigg)^{4/9}=\\& 6.6\times 10^{50}L_{52}^{6/5} \bigg(\frac{\lambda}{\epsilon}\bigg)_8^{-1/5} \eta_3^{-1}\mbox{ erg}/\mbox{sec}/\mbox{sterad},
\end{split}
\end{equation}
and the corresponding temperature is
\begin{equation}
\label{eq:Tph}
T_{ph}=T_{ph}'\Gamma(r_{ph})=110 L_{52}^{3/10} \eta_3^{1/4} \bigg(\frac{\lambda}{\epsilon}\bigg)_8^{-11/20} \mbox{keV}.
\end{equation}
The temperature at the photosphere thus naturally resides close to the observed peak of the prompt GRB emission
This calculation assumes a pure black body spectrum. If photons are not produced efficiently enough, the temperature can get higher and a Wien Spectrum is obtained instead of a Planck spectrum \citep{Begue2015}. However, electrons accelerated due to reconnection below the photosphere may produce sufficient synchrotron photons for complete thermalization to take place.
A deviation from a black-body spectrum, could also occur due to IC scatterings of the thermal photons by the relativistic electrons in the flow. This could build-up a high energy tail above the thermal peak. However as shown in \S \ref{sec:Comptonization} the effective Compton $Y$ parameter in this model is expected to be much smaller than unity, $Y\lesssim 0.01$, implying that although IC effects may distort somewhat the shape of the blackbody spectrum, their effects are energetically sub-dominant and cannot significantly change the typical temperature and flux.

For $r_{ph}<r<r_s$, matter and radiation decouple. The resulting emission is non-thermal and since the jet is highly magnetized, it will be dominated by synchrotron emission \citep{Beniamini2014}.
The properties of the emission are characterized by $\nu_{syn}$, the synchrotron frequency emitted by the typical energy electrons (these are either electrons with $\gamma=\gamma_i$ for $p>2$ or electrons with $\gamma=\gamma_f$ for $p<2$) and $\nu_c$,
the synchrotron frequency emitted by electrons that cool over the dynamical time-scale.
\begin{equation}
\begin{split}
\label{eq:nusyn}
& \nu_{syn}=\Gamma \gamma_e^2 \frac{qB'}{2\pi m_e c}\approx
2 \times 10^{22}\bigg(\frac{r_{ph}}{r}\bigg)^{5/3}\frac{\eta_3^3}{L_{52}^{1/2}}\bigg(\frac{\epsilon_e}{\xi}\bigg)^{2}\mbox{Hz}\\
&=1.4 \times 10^{19}\bigg(\frac{r_s}{r}\bigg)^{5/3}\frac{L_{52}^{1/2}}{\eta_3^{2}} \bigg(\frac{\epsilon_e}{\xi}\bigg)^2 \bigg(\frac{\lambda}{\epsilon}\bigg)_8^{-1} \mbox{Hz}
\end{split}
\end{equation}
\begin{equation}
\begin{split}
\label{eq:nuc}
& \nu_c=\frac{72 \pi q m_e c^3 \Gamma^3}{\sigma_T^2B'^3 r^2}\approx 4 \times 10^8\bigg(\frac{r}{r_{ph}}\bigg)^{3}\frac{L_{52}^{3/10}}{\eta_3}\bigg(\frac{\lambda}{\epsilon}\bigg)_8^{-4/5}\mbox{Hz}\\
&=10^{16}\bigg(\frac{r}{r_s}\bigg)^{3} \frac{\eta_3^{8}}{L_{52}^{3/2}}\bigg(\frac{\lambda}{\epsilon}\bigg)_8 \mbox{Hz}
\end{split}
\end{equation}
where $q$ is the electron charge and $\sigma_T$ is the Thomson cross section. If $\eta_3\gtrsim 3$, then $\sigma(r_{ph})\gtrsim 10$ and $p(r_{ph})\lesssim 2$. In this case, at the photosphere, $\nu_m=\nu(\gamma_f)$. For smaller values of $\eta$, $\nu_{syn}=\nu(\gamma_i)$ (and since $\sigma$ decreases with $r$, this remains the case up until $r_s$). The numerical expression in Eq. \ref{eq:nusyn} tends to over predict the value of $\nu_{syn}(r_{ph})$ when $2\lesssim \eta_3 \lesssim 6$ and the factor $((p-2)/(p-1))^2$ deviates significantly from unity.
The values of $\nu_c$ close to the photosphere, as given by Eq. \ref{eq:nuc}, are extremely low, and imply that electrons would very quickly cool down to extremely low values of $\gamma \beta \approx 1$ and below (recall that $\gamma_c \beta_c=\gamma_e \beta_e (\nu_c/\nu_{syn})^{1/2}$). In this regime, the regular synchrotron expressions do not hold any longer, and as a result, the expression for $\nu_c$ should be modified. However, as shown below $\nu_{SSA}(r_{ph})\gg \nu_c(r_{ph})$. Therefore, in any case, the synchrotron emission becomes self absorbed for electrons with
$\gamma<\gamma_{SSA}$ and they do not reach $\gamma_c$ within a dynamical time.

For $r\gtrsim r_{ph}$ we obtain that $\nu_{syn}\gg \nu_c$, i.e. the synchrotron radiation is strongly in the ``fast cooling" regime \citep{Beniamini2014}. This provides further justification to the simplifying assumption that the thermal and non-thermal emission are
largely decoupled as particles cool via synchrotron on a time-scale much shorter than the dynamical one, whereas IC losses are suppressed due to the Klein-Nishina effect. Furthermore, we note that $\nu_{syn}(r_{ph})$ typically resides in the gamma-rays.
Since $\nu_m$ decreases with radius, while $\nu_c$ increases with radius at an even greater pace, it is possible that $\nu_c$ becomes larger than $\nu_m$ before $r_s$.
This occurs for
\begin{equation}
\label{eq:rtrans}
\eta_3>1.8 L_{52}^{1/5}\bigg(\frac{\epsilon_e}{\xi}\bigg)^{1/5}\bigg(\frac{\lambda}{\epsilon}\bigg)_8^{-1/5}.
\end{equation}
The corresponding crossover frequency is 
\begin{equation}
\label{eq:nutrans}
\nu=2\times 10^{18}\frac{\eta_3^{11/7}}{L_{52}^{3/14}}\bigg(\frac{\epsilon_e}{\xi}\bigg)^{9/7}\bigg(\frac{\lambda}{\epsilon}\bigg)_8^{-2/7}\mbox{Hz}.
\end{equation}

For ``fast cooling" (which as noted above is the case for the range of radii that typically dominates the emission) the maximal synchrotron spectral luminosity (per sterad), at $\nu_c$, is given by \citep{Sari1998}:
\begin{equation}
\begin{split}
&L_{\nu,max}=\frac{m_e c^2 \sigma_T \Gamma B' N_e}{3q}\\&=5\times 10^{34} \frac{L_{52}^{1.3}}{\eta_3^2}\bigg(\frac{\xi}{0.2}\bigg) \bigg(\frac{\lambda}{\epsilon}\bigg)_8^{1/5} \mbox{erg Hz}^{-1}\mbox{sec}^{-1}
\end{split}
\end{equation}
where $N_e$ is the number of particles in a causally connected width at a radius $r$ (and within a solid angle of 1 sterad), $N_e=r\dot{N}_e(r)/(2c\Gamma^2)$. One can arrive at the same result by assuming that a factor $\epsilon_e/2$ of the power released in the range $[r,r+dr]$, $d\dot{E}(r)$, is radiated via the synchrotron process at $\nu_m$ (assuming $\nu_m>\nu_c$), as is indeed expected in the fast cooling regime.
For other frequencies, one has:
\begin{equation}
\label{eq:Lnu}
L_{\nu}=\left\{ \!
  \begin{array}{l}
L_{\nu,max}(\nu/\nu_c)^{1/3} \quad  \nu_c>\nu\\
L_{\nu,max} (\nu/\nu_c)^{-1/2} \quad \nu_{syn}>\nu>\nu_c\\
L_{\nu,max} (\nu_{syn}/\nu_c)^{-1/2}(\nu/\nu_{syn})^{-p/2} \quad \nu>\nu_{syn}\\
  \end{array} \right.
\end{equation}
This of course holds only so long as the spectrum does not become self absorbed, which happens for $\nu=\nu_{SSA}$, where:
\begin{equation}
\label{eq:nuSSA}
\frac{2 \nu_{SSA}^2}{c^2}\gamma(\nu_{SSA})\Gamma m_e c^2 \frac{\pi R^2}{\Gamma^2}=L_{\nu}(\nu_{SSA})
\end{equation}
where $\gamma(\nu_{SSA})$ is the Lorentz factor of electrons radiating synchrotron at a typical frequency $\nu_{SSA}$. For typical parameters, $\nu_c<\nu_{SSA}<\nu_m$ which using Eqs. \ref{eq:Lnu}, \ref{eq:nuSSA} results in:
\begin{equation}
\nu_{SSA}=1.5\times10^{18} L_{52}^{2/15} \bigg(\frac{\xi}{0.2}\bigg)^{1/3}\bigg(\frac{r}{r_{ph}}\bigg)^{-4/9}\bigg(\frac{\lambda}{\epsilon}\bigg)_8^{-7/5}\mbox{Hz}.
\end{equation}
Below this frequency, the spectrum becomes self absorbed.
Since the electrons are still fast cooling in this regime, the spectral slope below $\nu_{SSA}$ is $F_{\nu}\propto \nu^{11/8}$ \citep{GSP2000,GS2002}.
This can naturally provide a significant reduction of the optical and X-ray fluxes, as is required in order not to overproduce the flux in these bands as compared with the upper limits from observations \citep{Beniamini2014}. Rarer cases, in which the optical flux is apparently correlated with the $\gamma$-rays, e.g. GRB 041219B \citep{Vestrand2005} or GRB 080319B \citep{Racusin2008,Beskin2010}, may be the result of a combination of physical parameters (such as larger $\lambda/\epsilon,\eta,L$ or smaller $\xi$ or distance) that corroborate to sufficiently increase $F_{\nu}(\nu<\nu_{SSA})$.
A more quantitative examination would of course require physical knowledge of the underlying parameter distributions.

The non-thermal synchrotron photons could lead to the creation of a significant amount of pairs. This leads to a cut-off in the spectrum at some frequency $\nu_{co}$. Additionally, the newly formed pairs result in an increase of the particle density and hence of the photospheric
radius as compared with Eq. \ref{eq:rphot}. The optical depth for pair creation is given by:
\begin{equation}
\tau_{\gamma \gamma}(\nu,r)= \int_r^\infty \frac{11}{180}\Gamma(1-\beta)\sigma_{T} \frac{d_L^2 \int_{\nu_{an}}^{\infty} N_{\nu}d\nu}{r^2\Gamma c} dr
\end{equation}
where $\sigma_T$ is the Thomson cross section, $\nu_{an}=\Gamma^2 m_e^2 c^4/(h^2 \nu)$ is the frequency of a photon that can annihilate a photon with frequency $\nu$ and $d_L^2 \int_{\nu_{an}}^{\infty} N_{\nu}d\nu/(r^2\Gamma c)$ is the co-moving number density of annihilating photons.
The pair creation cut-off frequency satisfies the condition $\tau_{\gamma \gamma}(\nu_{co})=1$.
The creation rate of positrons + electrons is then
\begin{equation}
\dot{N}_{pairs}=2d_L^2 \int_{\nu_{co}}^{\infty}N_{\nu} d\nu.
\end{equation}
In order to significantly change the photospheric radius, this rate must be larger than $\dot{N}$.
Since the creation rate depends on the number of photons in the high energy portion of the spectrum and the latter depends exponentially on $p(\sigma)$, one has to numerically estimate $\nu_{co}$ and the change in $r_{ph}$ due to pairs. For our canonical choice of parameters in this work, given in Fig. \ref{fig:spectra}, $h\nu_{co}\approx 500$MeV and the change in $r_{ph}$ as compared with Eq. \ref{eq:rphot} is negligible.
\section{Results}
\label{sec:results}
\subsection{The Spectrum}
The maximum angular time-scale is set at the saturation radius: $t\sim R/2 c \Gamma^2 \lesssim r_s/ 2 c \Gamma_{\infty}^2\simeq 0.1 \lambda/ (\epsilon c) \simeq 3 \times 10^{-4} (\lambda/\epsilon)_8$sec. 
This time is much shorter than the typical duration of a single pulse in the light-curve. Therefore, within this model, any temporal behaviour in the GRB light-curve reflects directly the activity of the engine. This is generic to all GRB models that involve emission from the surroundings of the photosphere.
Since we are considering a continuous energy injection, this implies that the observer receives the integrated radiation emitted released between $r_{ph}$ and $r_s$.

\begin{figure*}
\centering
\includegraphics[scale=0.35]{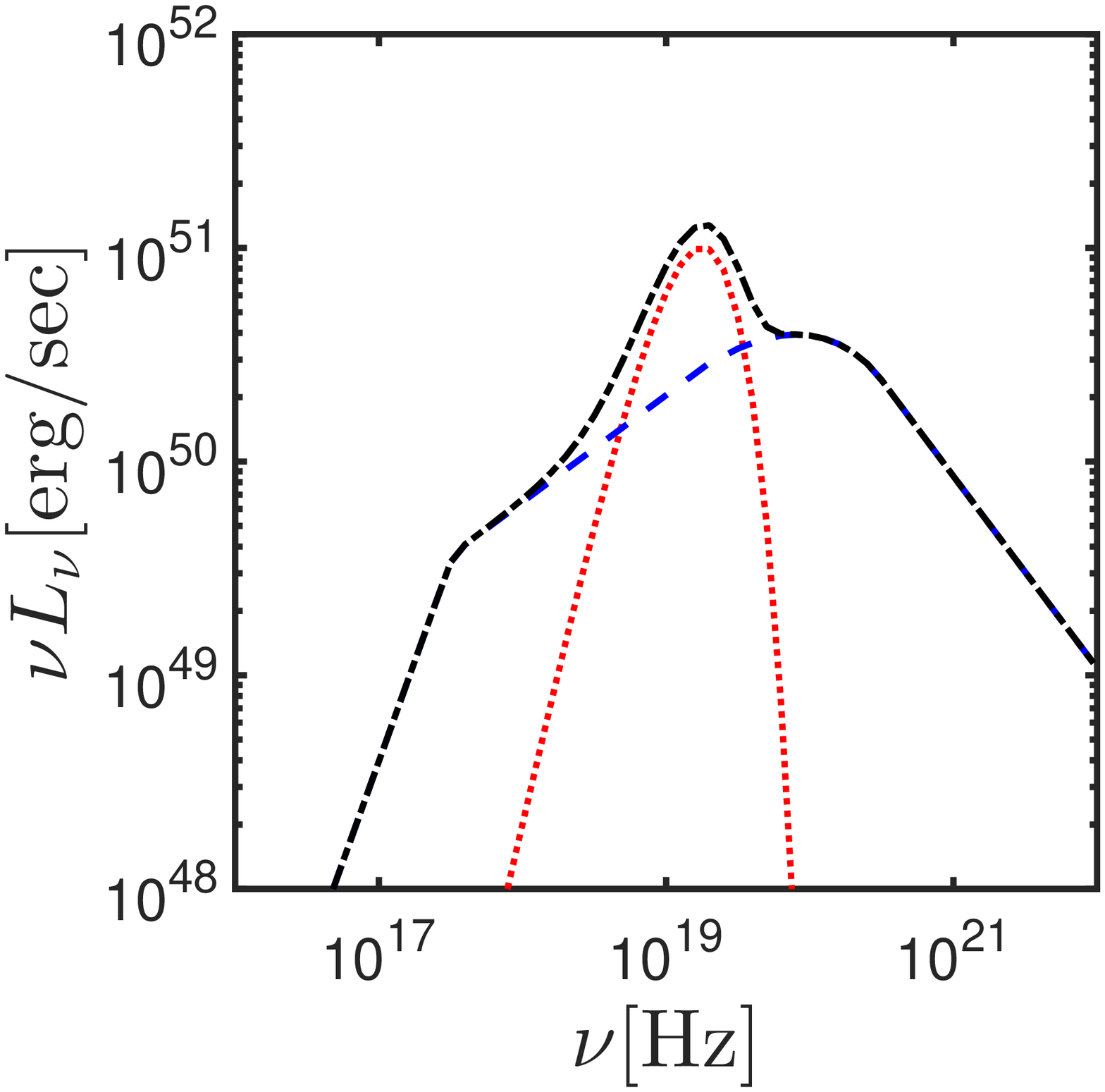}
\includegraphics[scale=0.35]{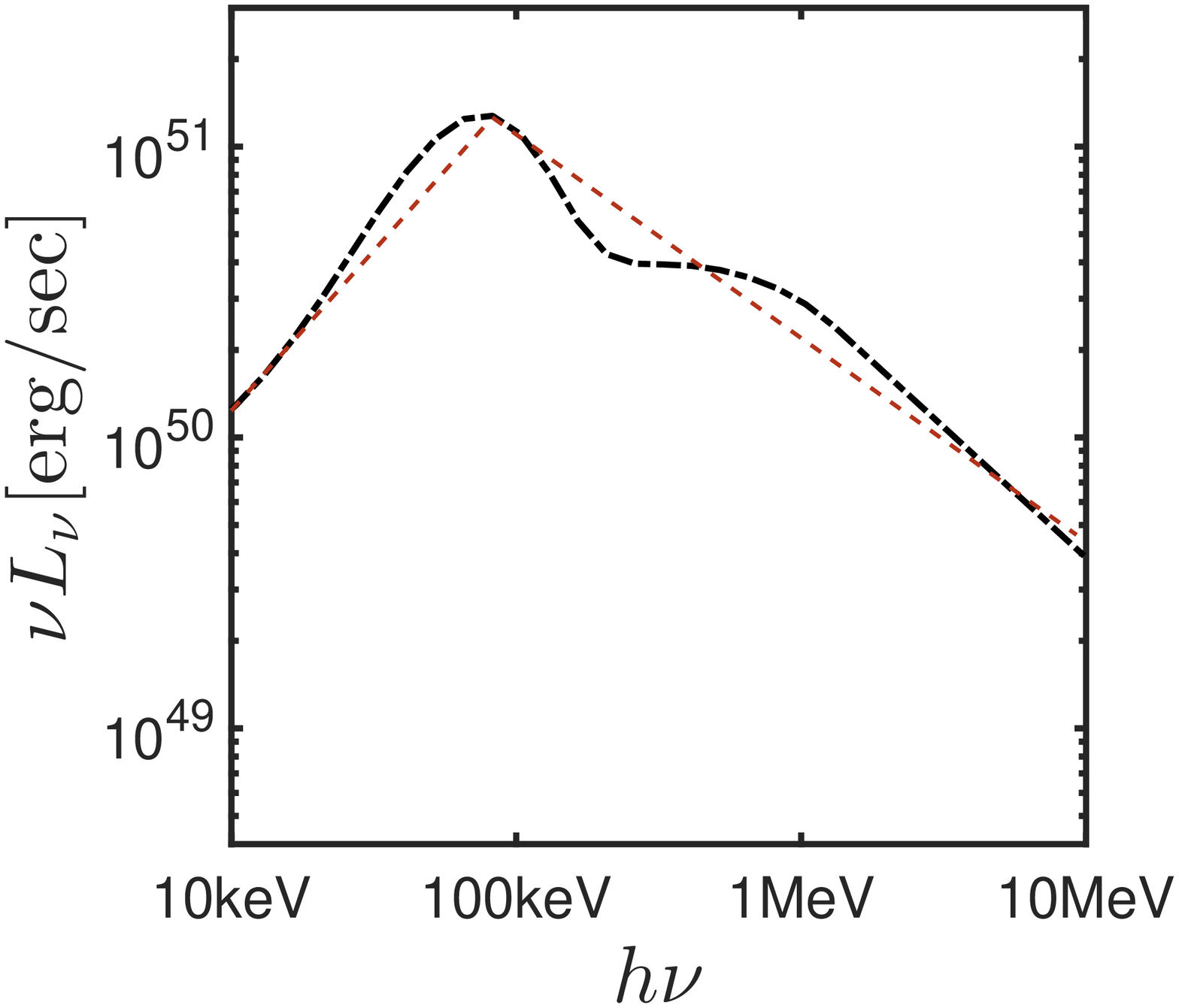}
\caption
{\small Spectrum obtained for: $\eta_3=1/3,(\lambda/\epsilon)_8=4,L_{52}=1,\xi=0.2$.
Left: The dashed line is the spectrum from the non-thermal components emitted at $r_{ph}<r<r_s$ (gradually shifting to smaller frequencies as the radius increases). A dotted line marks the narrow component (assumed for simplicity to be thermal) emitted at $r_{ph}$.
Finally, the dot-dashed line indicates the overall (thermal + non-thermal) spectrum.
Right: Close up of the spectrum in the 10keV-10MeV range (typical for prompt $\gamma$-ray observations).
The low energy spectra approximately follows $\nu L_{\nu} \propto \nu^{1.1}$ whereas the high-energy part falls as: $\nu L_{\nu} \propto \nu^{-0.7}$.}
\label{fig:spectra}
\end{figure*}

In fig. \ref{fig:spectra} we plot the resulting spectra obtained from a steady injection of energy into the jet.
For the purposes of illustration, we assume here $\eta_3=1/3,(\lambda/\epsilon)_8=4,L_{52}=1,\xi=0.2$ and a typical red-shift $z=1$.
Since $\sigma$ is already quite low at the photosphere, $\sigma(r_{ph})\approx 1.3$, the emission is dominated by the thermal component, which peaks at $\approx 100$keV, consistent with prompt GRB observations \citep{Kaneko2006,Nava2011}.
At the same location, the synchrotron emission peaks at $\approx$2MeV. As the radius increases the synchrotron emission is shifted to lower frequencies. This evolution is mainly responsible for building the high energy spectral slope of the $\gamma$-rays, and to a lesser extent also softens somewhat the low energy spectrum. Below $\approx 2$keV the emission becomes strongly suppressed due to SSA.
The spectrum between $\approx 10$keV (typical for the low end cut-off of $\gamma$-ray detectors) and the peak is reasonably described by a power law: $\nu L_{\nu} \propto \nu^{1.1}$ (or $\alpha=-0.9$ where $N_{\nu} \propto \nu^{-\alpha}$ for $\nu<\nu_p$). This is consistent with the typical values found in observations \citep{Kaneko2006,Nava2011}, which are not easily accounted for by most GRB models. 
The high energy part is less smooth, due to the fact that the thermal spectrum falls exponentially in this regime. On average, the spectrum between the peak frequency and 10MeV (typical for the high end cut-off for $\gamma$-ray detectors) can be fitted by a power law decline with a slope: $\nu L_{\nu} \propto \nu^{-0.7}$ ($\beta=-2.7$ where $N_{\nu} \propto \nu^{-\beta}$ for $\nu>\nu_p$). This is consistent with observed values for some prompt GRB spectra although slightly steeper than the average value.
However, due to the strong dependency of $\nu_{syn}(r)$ on $\xi$, changing, for instance, $\xi$ from $0.2$ to $0.1$ is sufficient to lead to a more typical $\beta=-2.35$. This has almost no effect on the peak frequency, and increases also the low energy spectral slope to $\alpha=-0.7$ (which is still very common).

\subsection{Predictions / observables of the model}
The efficiency of prompt GRBs, defined as $\eta_{\gamma}\equiv E_{\gamma}/(E_{\gamma}+E_{\rm kin})$ (where $E_{\rm kin}$ is the kinetic energy of the outflow before it decelerates due to interaction with the external medium) can be easily estimated within this model.
For fast cooling electrons, at any given radius, $dL_{syn}\approx d\dot{E}_e\approx (\epsilon_e/2)d\dot{E}$ (see \S \ref{sec:rad}).
Since $\int_{r_{ph}}^{r_s}d\dot{E}\approx (1/2)L$ (the exact value in the last expression depends of course on the location of the photosphere), the total synchrotron luminosity is given by $L_{syn}\approx (\epsilon_e/4)L$.
Combining this with the thermal luminosity (Eq. \ref{eq:Lph}) gives a total efficiency of $\eta_{\gamma} \approx 0.1$ for the canonical parameters considered here.
This is in accordance with estimates of the efficiency based on afterglow observations,
suggesting that $\eta_{\gamma}=0.1-0.2$ \citep{Beniamini2015,Beniamini2016}.

\begin{figure*}
\centering
\includegraphics[scale=0.3]{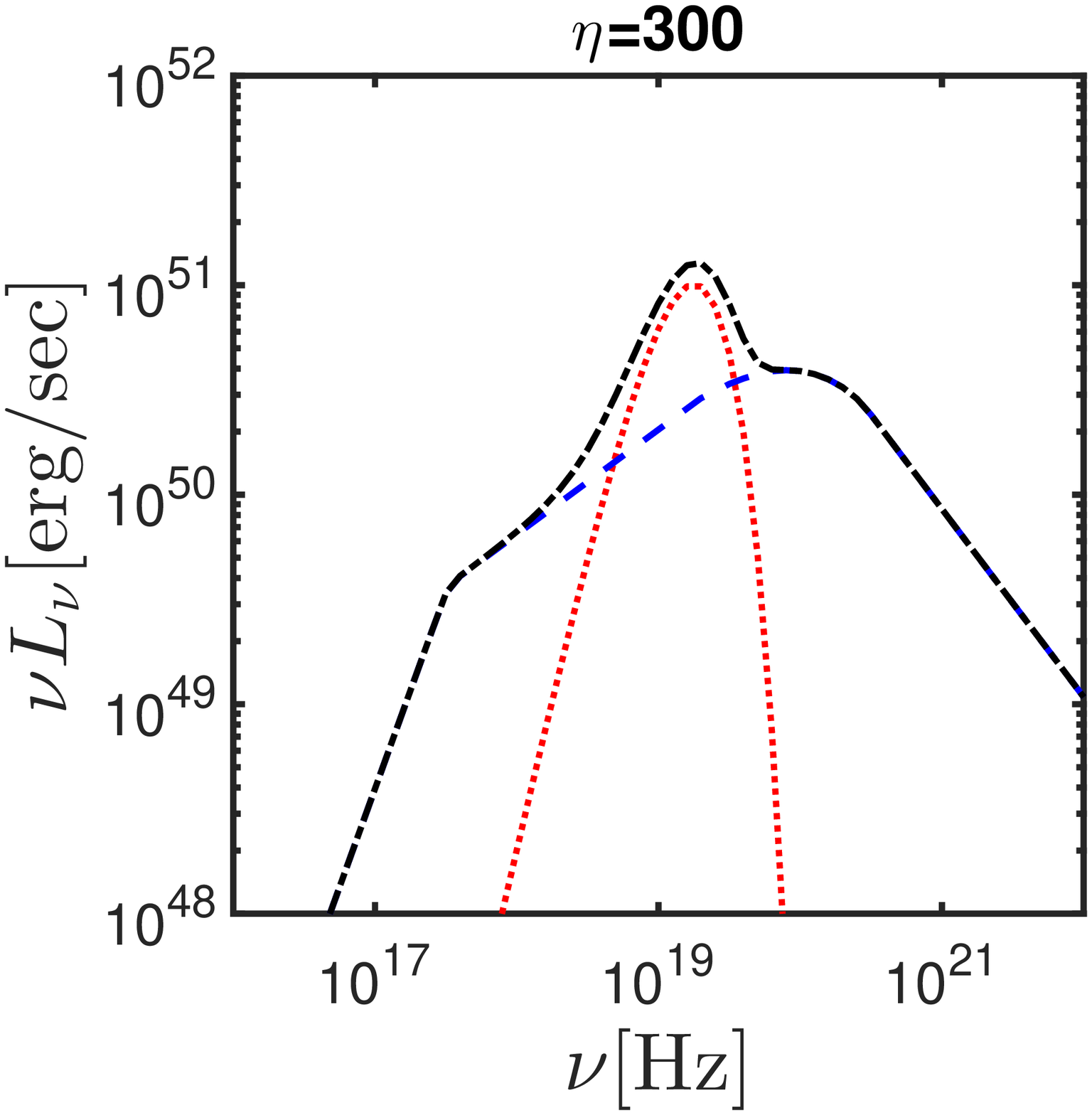}
\includegraphics[scale=0.3]{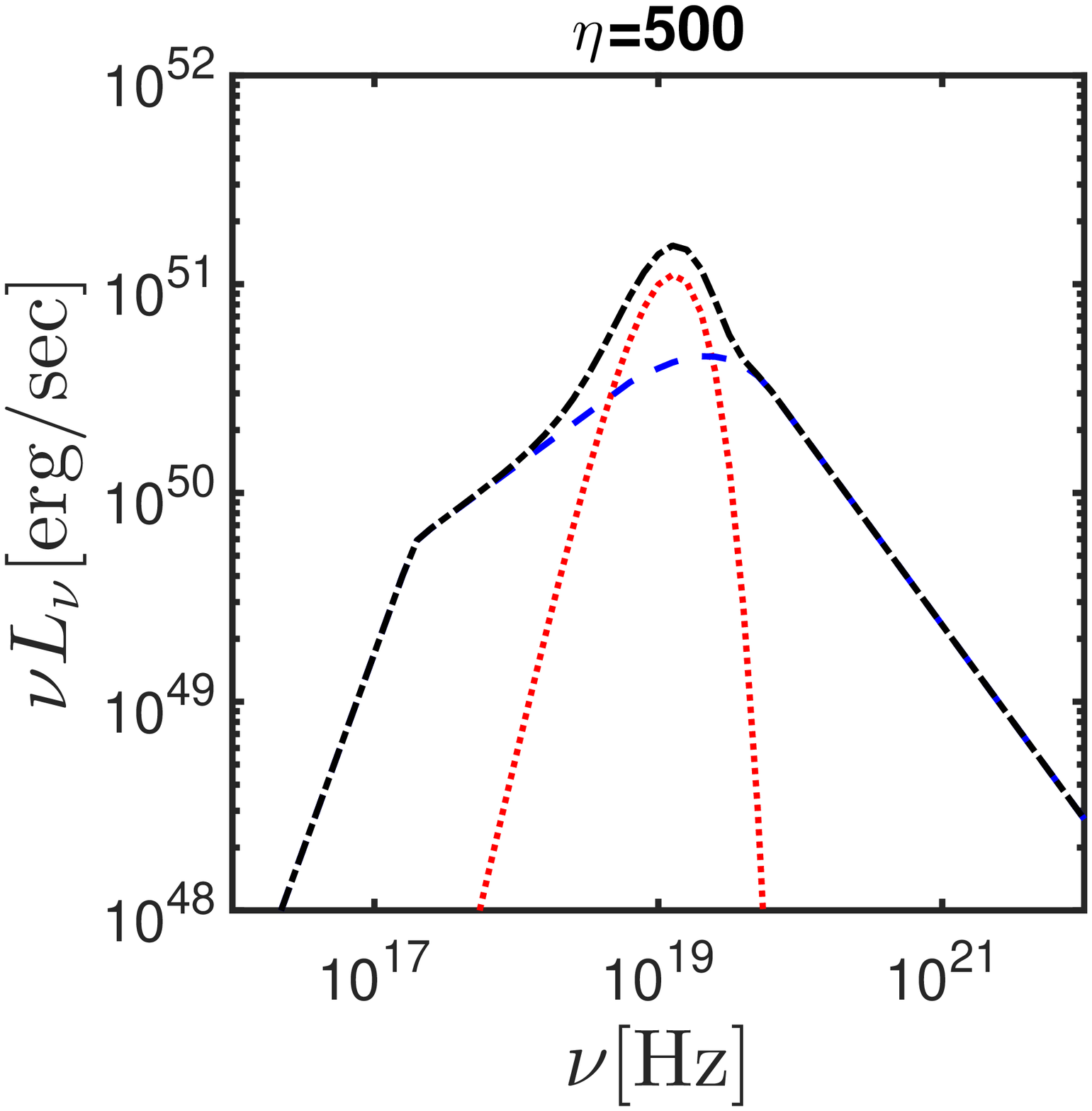}\\
\includegraphics[scale=0.3]{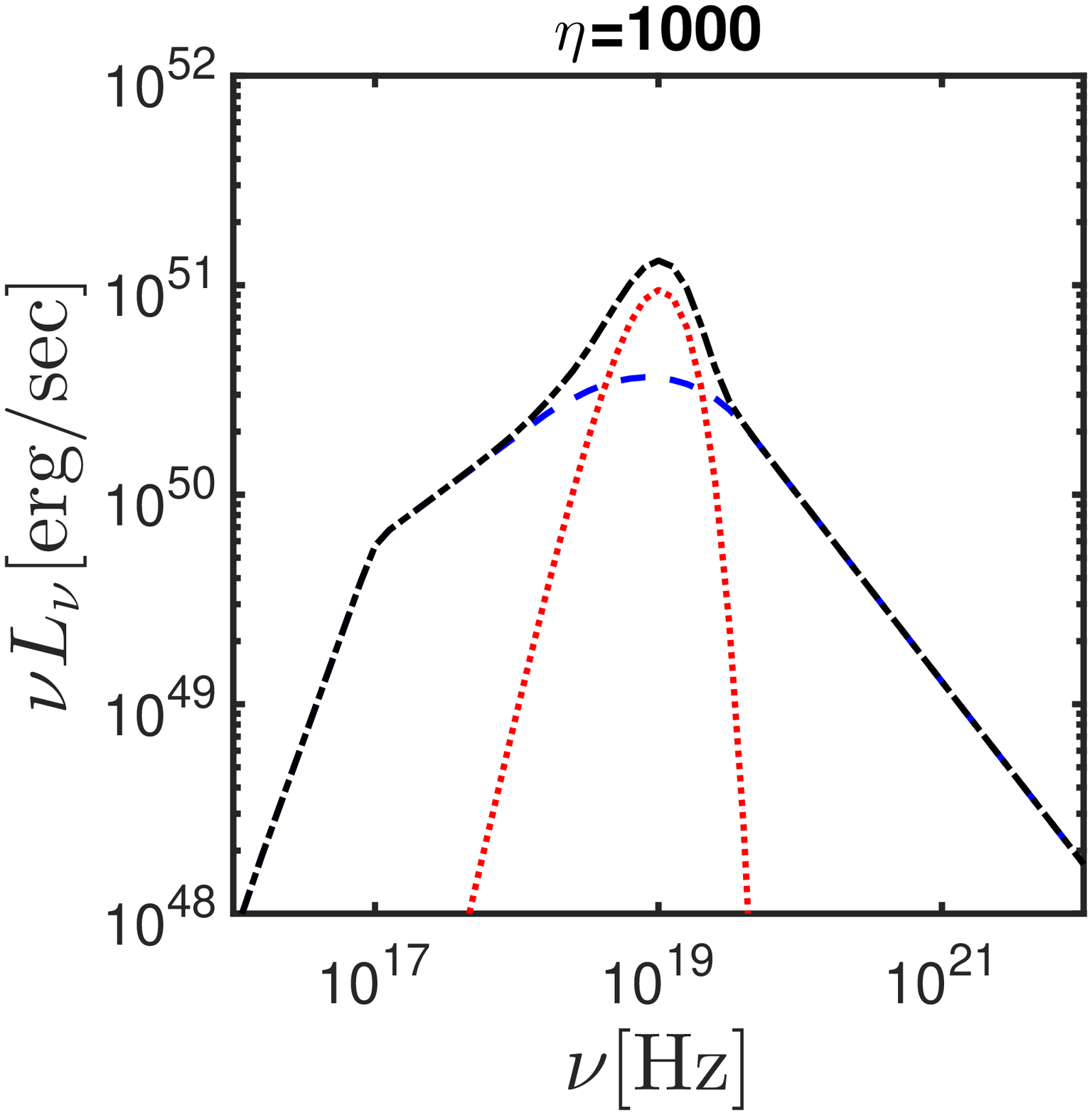}
\includegraphics[scale=0.3]{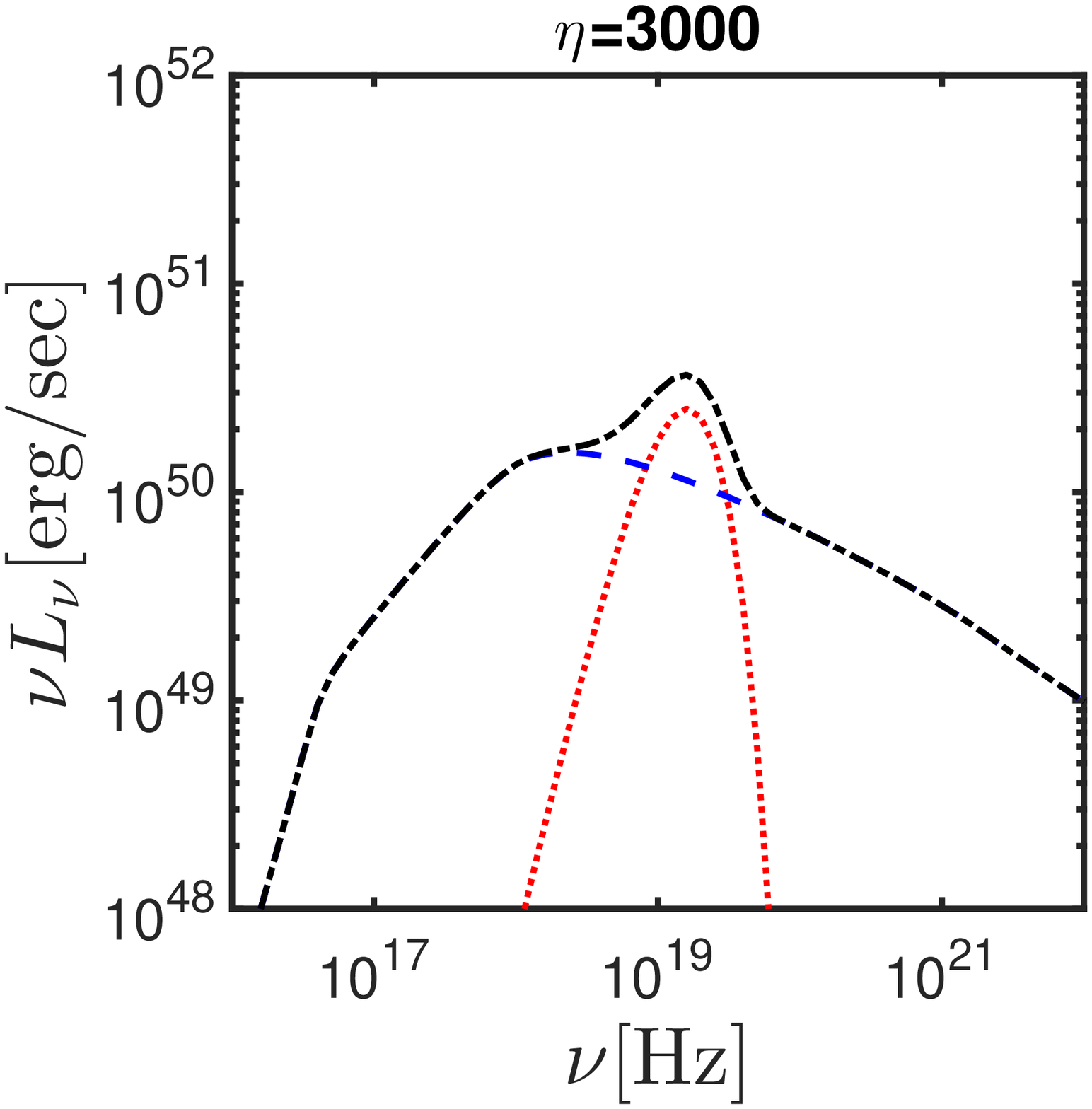}
\caption
{\small Spectra obtained for: $(\lambda/\epsilon)_8=4,L_{52}=1,\xi=0.2$ and different values of $\eta$. Starting from top left and in clock-wise order, $\eta=0.3,0.5,1,3$.}
\label{fig:eta}
\end{figure*}

The narrow photospheric component typically peaks at around a few hundred keV. This temperature is not so sensitive to the model parameters. However, the thermal luminosity (relative to the overall luminosity) decreases with $\eta$ (or equivalently $\sigma_0$) and is less sensitive to the other parameters, provided that $r_s>r_{ph}$ (see below for discussion of the other limit).
At the same time, the synchrotron flux changes significantly with $\eta$.
At the high-energy portion of the spectrum, there are competing effects. $\nu_{syn}(r_{ph})$ increases with $\eta$. However, the same trend also leads to a decrease in the magnetization
and the value of $p$. This results in more pair creation, which may become sufficient to push the photosphere outwards and decrease the typical synchrotron frequency.
The most important change, however, is at the low-energy end of the spectrum (down to a few keV).
This is typically controlled by $\nu_{syn}(r_s)$ which decreases as $\eta^{-2}$.
This leads to the observable prediction of the model according to which bursts with softer low energy spectral slopes have a weaker thermal bump.
Since $p$ decreases with $\eta$, the non-thermal component in the spectrum these bursts, above their peak, would also be flatter and would be observable up to larger frequencies (before cutting off at the pair-creation cut-off or at the maximal synchrotron frequency).
The slope of $N_{\nu,syn}$ above $\nu_{syn}(r_{ph})$ is approximately given by:
\begin{equation}
-\frac{p(r_{ph})+2}{2}=-1-1.4(\lambda/\epsilon)_8^{0.06}\frac{L_{52}^{0.06}}{\eta_3^{0.45}}.
\end{equation}
For sufficiently clean jets, $\eta_3>1.8 L_{52}^{1/5}(\epsilon_e/\xi)^{1/5}(\lambda/\epsilon)_8^{-1/5}$ (see Eq. \ref{eq:rtrans}), the overall spectrum becomes slow cooling before the saturation radius (and for a gradually larger extent of radii, as $\eta$ increases). This results in an overall reduction of the non-thermal synchrotron luminosity.
The dependence of the spectra on $\eta$ described above, is shown in Fig. \ref{fig:eta} while leaving the other parameters constant at their value shown in Fig. \ref{fig:spectra}.

For smaller values of $\eta$ the spectrum gradually becomes more dominated by the narrow photospheric component.
As shown in \S \ref{sec:dynamics}, any material that is launched with $\eta \lesssim 290 (\lambda/\epsilon)_8^{-1/5}L_{52}^{1/5}$ would lead to a completely thermal signature. This may have in fact been observed. In a recent study, \cite{BK2016}, claimed that X-ray flares must be self absorbed between the optical and X-ray bands implying a small radius of emission and Lorentz factor for the material producing them $R \lesssim 3\times 10^{14} \mbox{cm}, \Gamma \lesssim 20$. These flares could therefore arise from material which is ejected by the central engine with an initially smaller total energy per Baryon.
In addition, the luminosity of the thermal emission in this case is much weaker and peaks in the X-rays \citep{GianniosSpruit2007}.
A possibly related observation is the optical transient source PTF11agg observed by the PTF. Observations of this source favour a cosmological transient source, but fail to account for the simplest expectations from an off-axis GRB \citep{Cenko2013}. \cite{Cenko2013} argue that the most likely origin for this transient is then a ``dirty fireball", which is a burst with significantly suppressed high-energy emission, that would nonetheless still produce an afterglow as the material collides with the external medium. In the model considered in this paper, due to strong suppression of the prompt luminosity when $r_s<r_{ph}$, such a situation would naturally arise if the prompt material is ejected with low enough $\eta$.

\begin{figure*}
\centering
\includegraphics[scale=0.3]{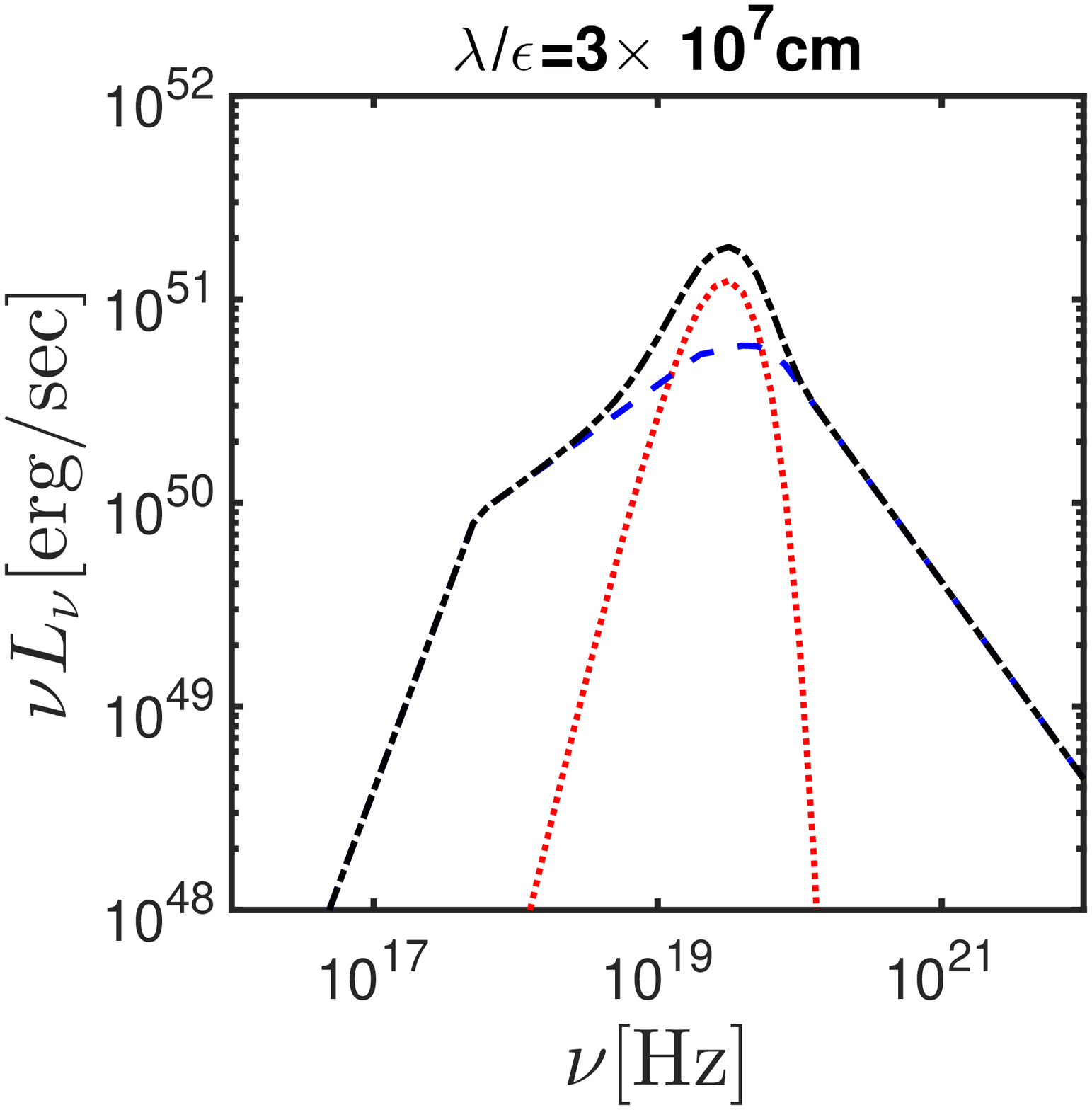}
\includegraphics[scale=0.3]{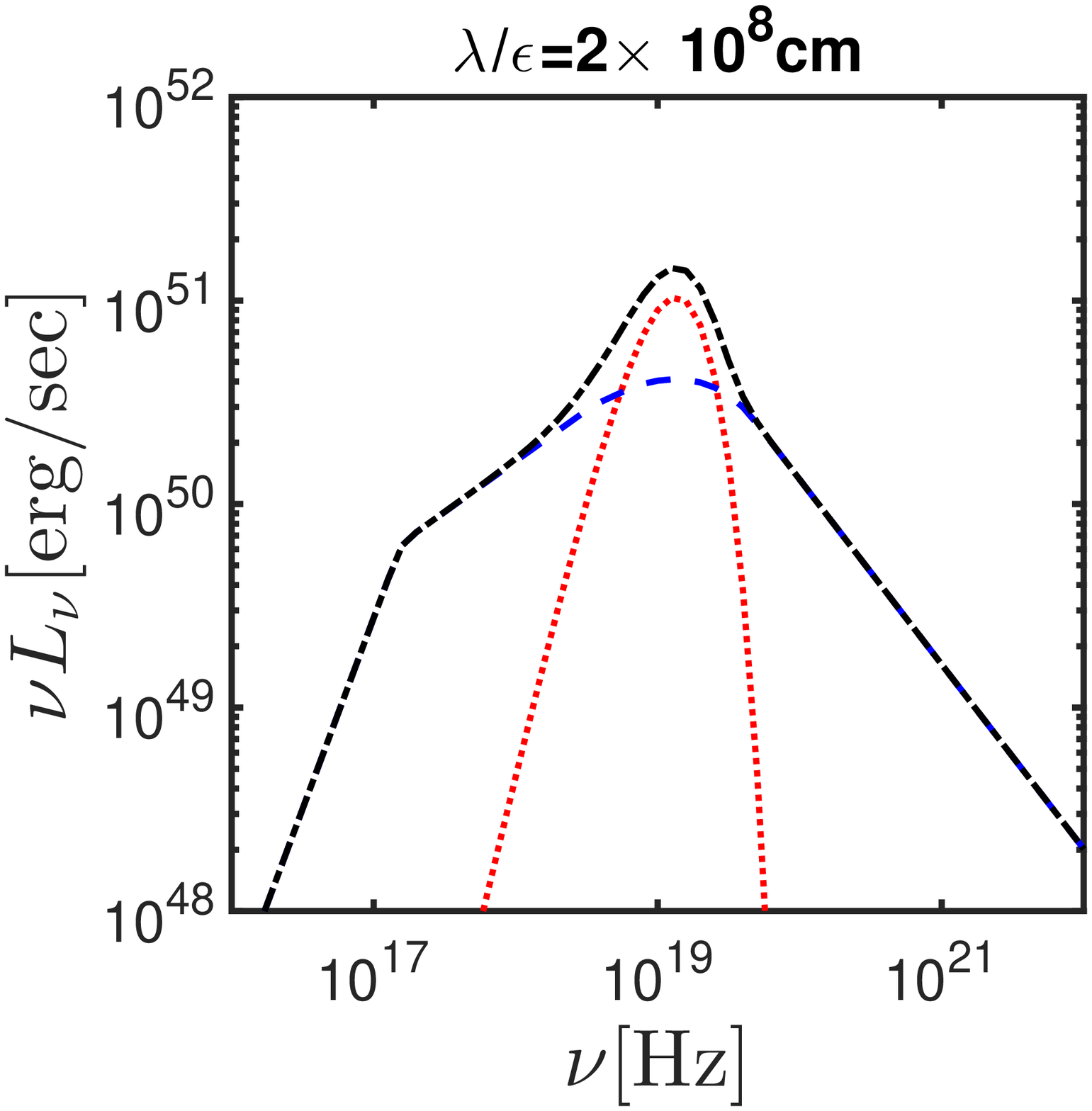}\\
\includegraphics[scale=0.3]{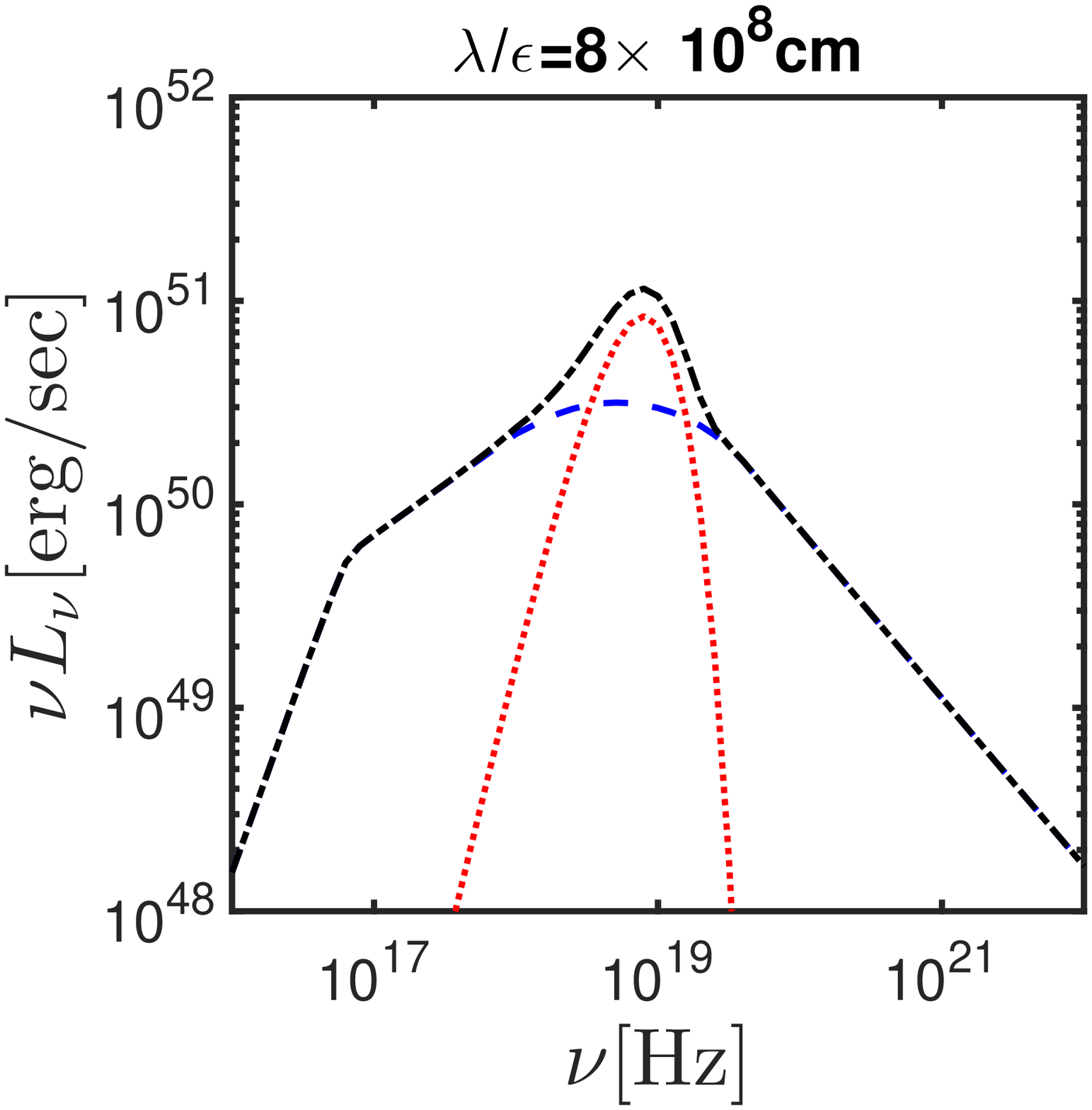}
\includegraphics[scale=0.3]{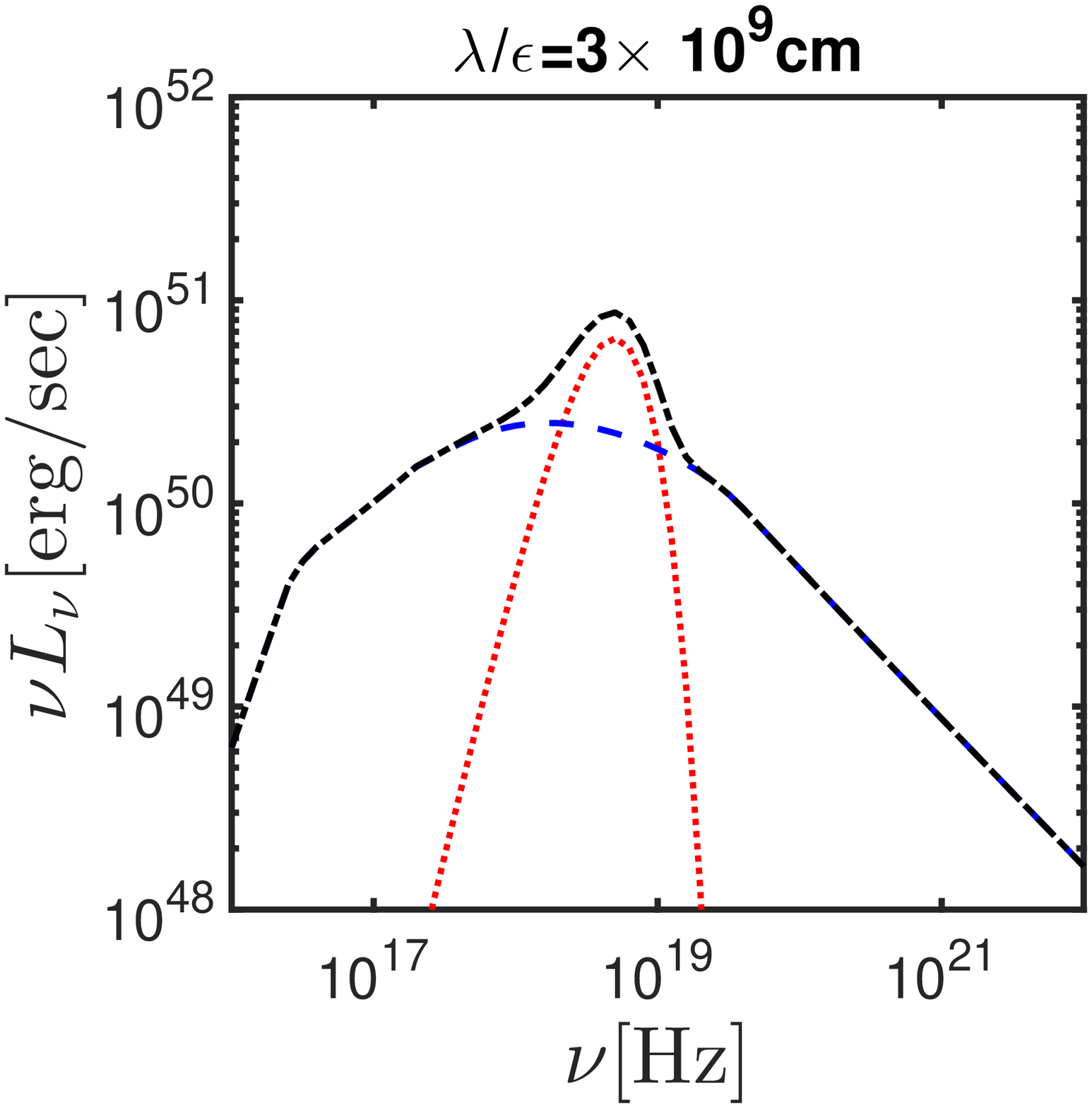}
\caption
{\small Spectra obtained for: $\eta_3=1,L_{52}=1,\xi=0.2$ and different values of $\lambda/\epsilon$. Starting from top left and in clock-wise order, $\lambda/\epsilon=3\times 10^7,2\times 10^8,8\times 10^8,3\times 10^9$cm.}
\label{fig:lambdaepsilon}
\end{figure*}

\begin{figure*}
\centering
\includegraphics[scale=0.3]{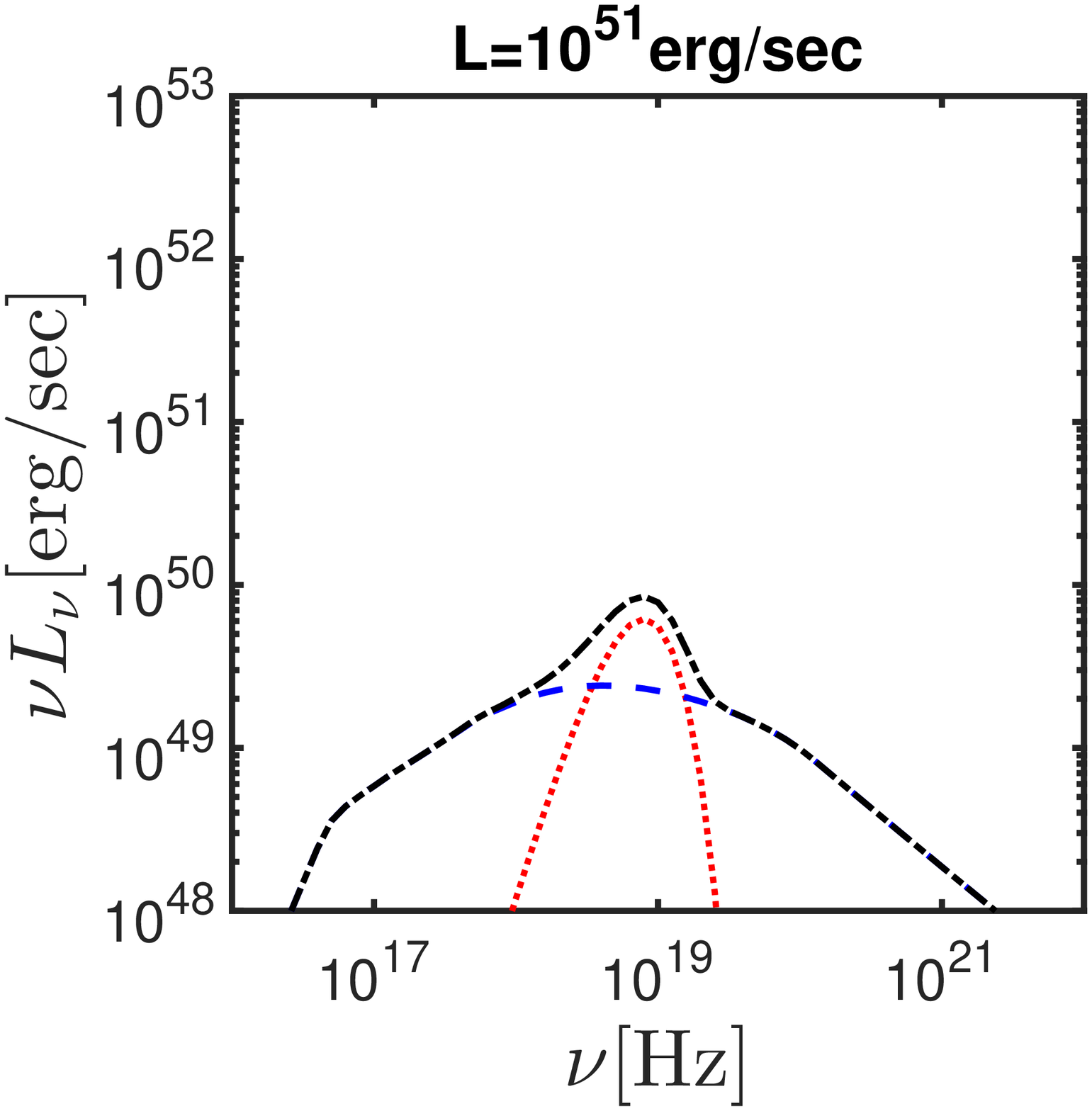}
\includegraphics[scale=0.3]{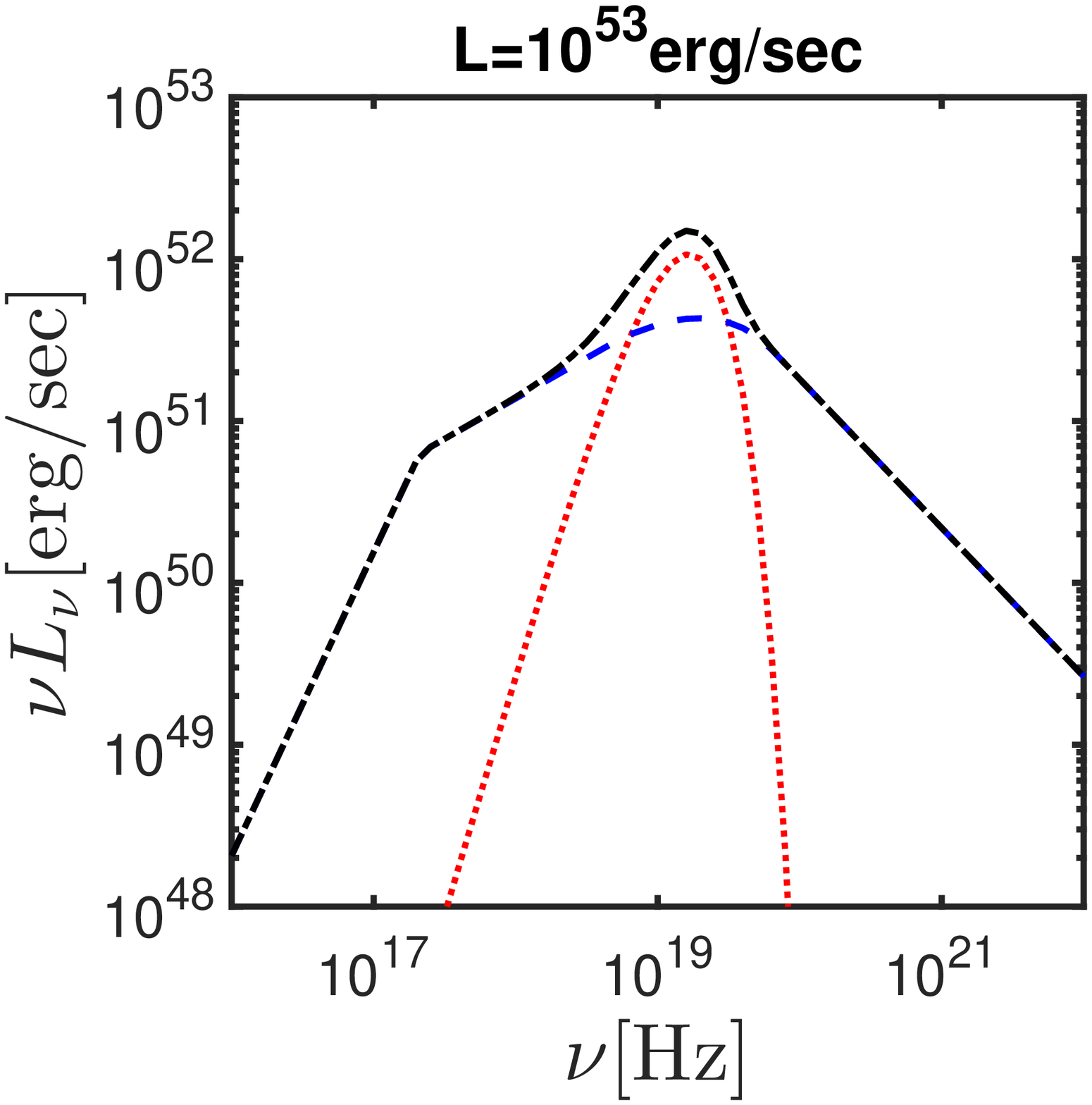}
\caption
{\small Spectra obtained for: $\eta_3=1,(\lambda/\epsilon)_8=4,\xi=0.2$ and different values of $L$. Left, $L_{52}=0.1$ and right, $L_{52}=10$.}
\label{fig:L}
\end{figure*}

\begin{figure*}
\centering
\includegraphics[scale=0.3]{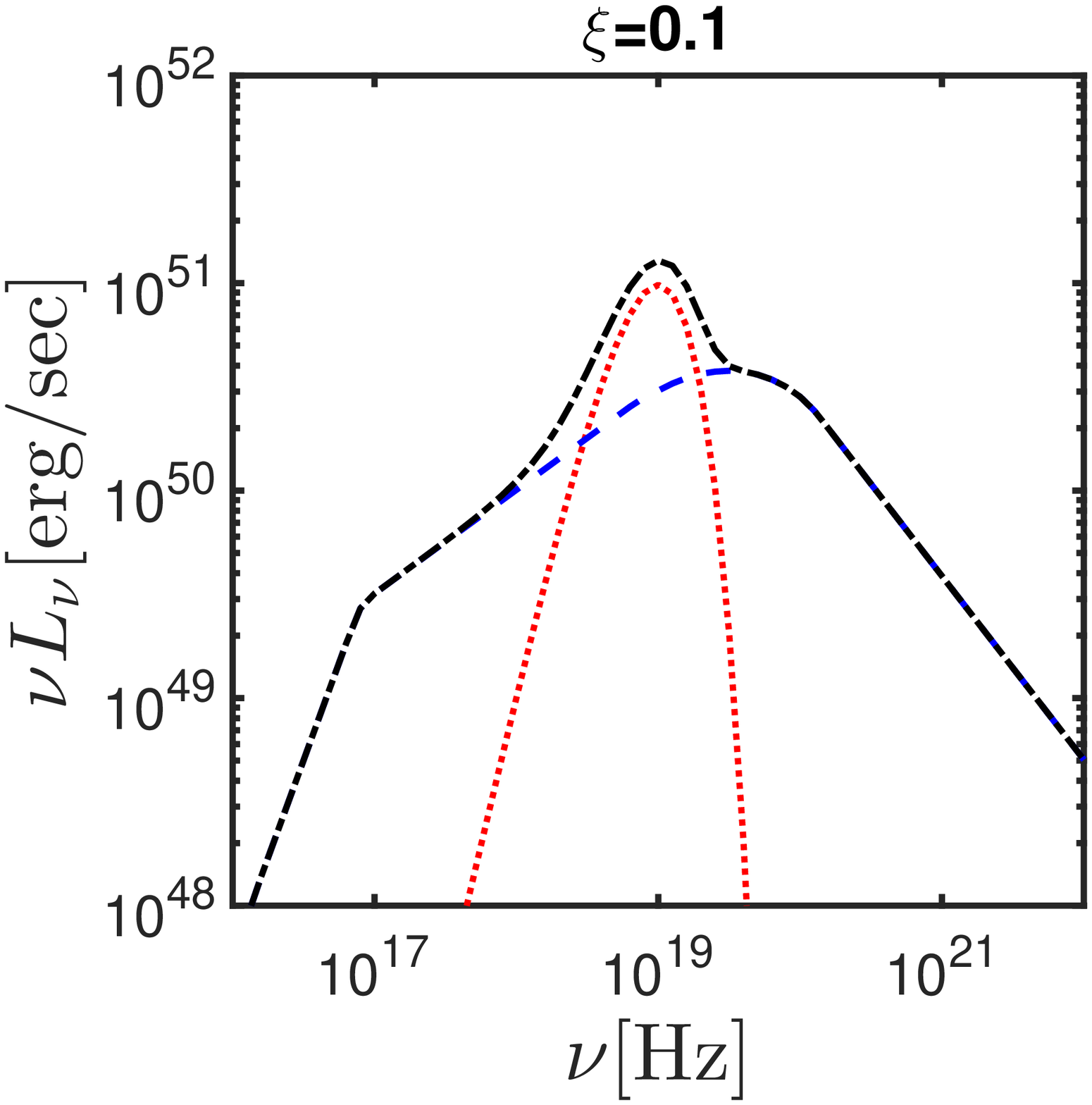}
\includegraphics[scale=0.3]{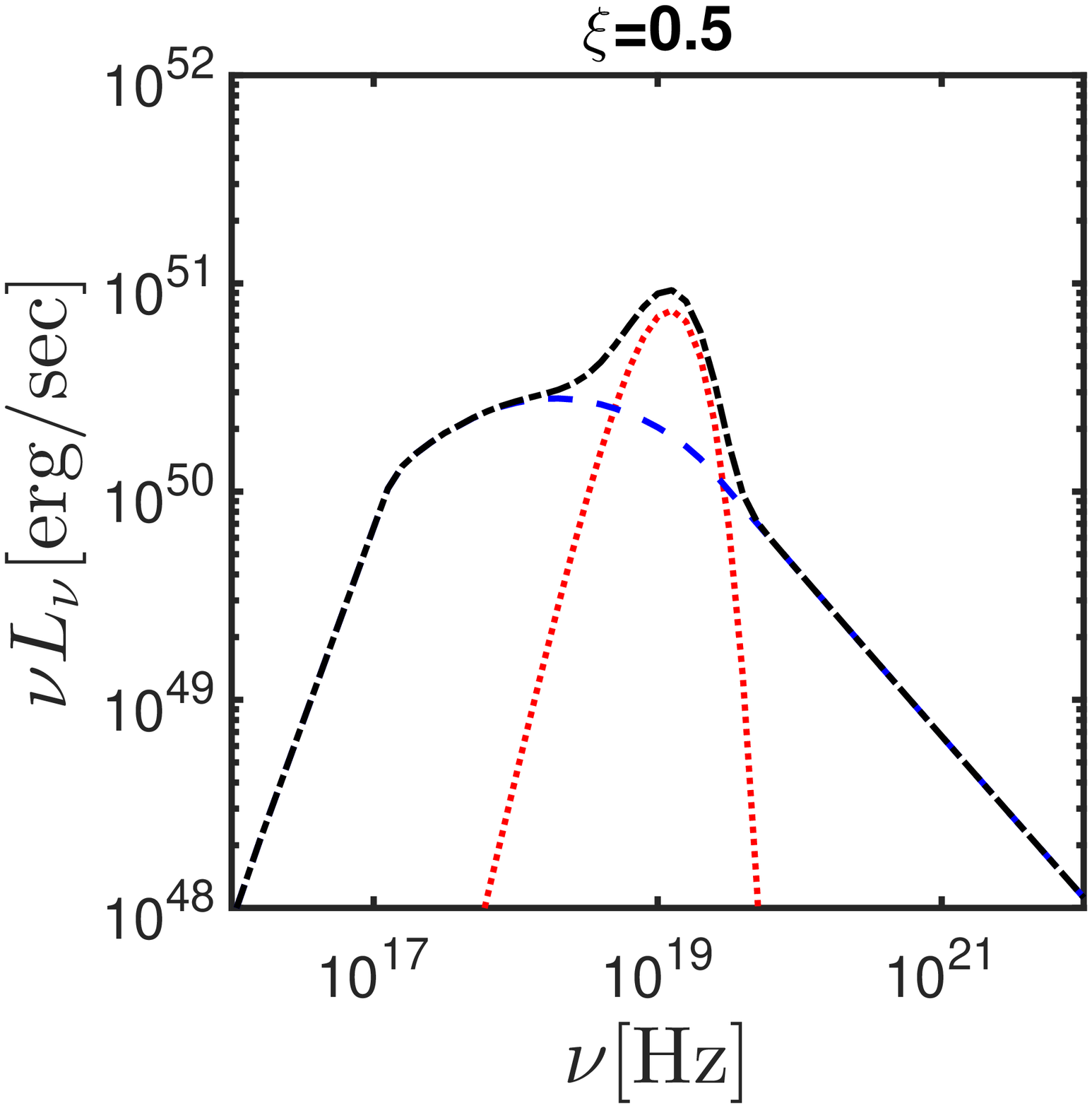}
\caption
{\small Spectra obtained for: $\eta_3=1,(\lambda/\epsilon)_8=4,L_{52}=1$ and different values of $L$. Left, $\xi=0.1$ and right, $\xi=0.5$.}
\label{fig:xi}
\end{figure*}

The dependence of the model on the other three parameters $\lambda/\epsilon,L$ and $\xi$ is shown in Figs. \ref{fig:lambdaepsilon},\ref{fig:L},\ref{fig:xi}. The main effect of increasing $\lambda/\epsilon$ is to reduce $\nu_{syn}(r_{ph})$, thus slightly shifting the non-thermal spectrum to lower-frequencies. $\lambda/\epsilon$ also strongly affects the SSA frequency. When it is small, the X-ray emission is strongly suppressed by absorption, whereas larger values could lead to detectable X-ray (and possibly even optical) emission during the prompt. The dependence on $L$ is straightforward, corresponding mainly to a linear increase in the non-thermal emission and a slightly stronger increase in the photospheric component. Finally, increasing $\xi$ reduces both $\nu_{syn}(r_{ph})$ and $\nu_{syn}(r_{s})$, resulting in smaller effective values of $\alpha,\beta$.

As mentioned in \S \ref{sec:dynamics}, the value of $\Gamma$ at the photosphere is almost constant in this model. Since most of the emission originates from this radius (and in particular the highest energy emission), this condition can be related to the maximal frequency of a synchrotron photon that can be accounted for by this model:
\begin{equation}
\label{eq:nusynmax}
h\nu_{syn,Max}(r_{ph})=5.5  L_{52}^{1/5} \bigg(\frac{\lambda}{\epsilon}\bigg)_8^{-1/5} \mbox{GeV}
\end{equation}
in the central engine frame. As shown in \S \ref{sec:results}, a cut-off in the spectrum may be seen at even lower frequencies, due to pair creation. Still, Eq. \ref{eq:nusynmax} provides a rough upper limit on the observable synchrotron photons in this model.
Although a few higher energy photons have been observed in some bursts, it has been argued that they could originate from the external forward shock \citep{KBD2009,Gao2009,KBD2010,Ghisellini2010}.

Finally, various correlations of the type $E_p-E_{iso}, E_p-E_{\gamma}$ or $E_p-L_p$ have been reported in GRB literature \citep{Amati2002,Ghirlanda2004,Yonetoku2004}.
Typically, studies of this type have found that $E_p\propto L_p^{1/2-2/3}$. Assuming that the location and luminosity of the peak is dominated by the thermal component, we have $T_{ph} \propto L_{ph}^{1/4} (\lambda/\epsilon)^{-1/2} \eta^{1/2}$. Therefore, a correlation similar to the observed one can be realized if, for instance,  $\eta \propto L^{\sim 1/2-5/6}$ and $\lambda/\epsilon$ is not strongly correlated with the other parameters (see also \citealt{GianniosSpruit2007}). Assuming this relation between $\eta$ and $L$, would imply that the above correlations would also hold for different pulses within a given burst, as was indeed observed in several cases \citep{Guiriec2015}. Furthermore, it would naturally result in an intensity tracking evolution within a given pulse \citep{Lu2012} and in pulse widening at lower frequencies \citep{Fenimore1995} and spectral lags \citep{Norris1996}.

\section{Summary}
We considered a model for the prompt phase of GRB emission arising from a magnetized outflow undergoing gradual energy dissipation due to magnetic reconnection. The dissipated magnetic energy is translated to bulk kinetic energy in the jet and to acceleration of particles. The energy stored in these particles is released in the form of synchrotron radiation as they gyrate around the strong magnetic fields in the jet. At small radii, the optical depth is large, and the emission is reprocessed through Comptonization into a narrow, quasi thermal, component. At larger distances the optical depth becomes small and the radiation escapes the system in non-thermal form.
The overall efficiency of the radiation (as compared with the jet power) is of order $0.1$, as required by afterglow observations.

The peak of the spectrum in this model is dominated by the photospheric component, and is expected to reside at the sub-MeV range, as observed in prompt GRBs. Furthermore, possible correlations with the luminosity (both when comparing different pulses in a given GRB, and between different GRBs) are possible depending on the inter-relations of the model's intrinsic parameters, the jet luminosity, $L$, the ``wavelength" of the structured magnetic field in the jet over the outflow velocity from the reconnection sites, $\lambda/\epsilon$ and the initial total energy per Baryon, $\eta$. This would then also lead to an intensity tracking evolution within a given GRB pulse, to pulse widening at lower frequencies and to spectral lags, all of which have also been reported in observational prompt GRB studies. For matter ejected from the central engine with $\eta \lesssim 290 (\lambda/\epsilon)_8^{-1/5}L_{52}^{1/5}$, the emission is completely thermalized. This may account for the common phenomena of X-ray flares. The temperature can be reduced to $\approx 4$keV, by assuming, for instance, $L=10^{52}\mbox{erg s}^{-1} (\lambda/\epsilon)=10^8\mbox{cm}, \eta=20$ \citep{GianniosSpruit2007}. This leads to a flare luminosity which is $L=6\times 10^{49}\mbox{erg s}^{-1}$ which is rather typical
for a flare occurring a few hundred seconds after the trigger \citep{Margutti2011}.
Due to the typically small signal to noise ratio, the spectra of these flares cannot be measured except for very few cases.
There is some suggestion that the brightest flares can be fitted (in the X-ray band) with a power-law spectrum: $F_{\nu}\propto \nu^{-1.1}$ \citep{Chincarini2010}, which would be incompatible with the expectations for the low $\eta$ material.
However the spectrum must become significantly harder at lower frequencies to be compatible with contemporaneous upper limits on the optical flux, which is suggestive of a photospheric origin \citep{BK2016}.

The non-thermal synchrotron component leads to a softening of the spectrum below the thermal peak and to hardening of the spectrum above it.
For our canonical parameters (see \S \ref{sec:results}), we find that the low energy spectral slope is $\alpha=-0.9$ and the high energy spectral slope is $\beta=-2.7$.
More generally, given the expected ranges of the intrinsic parameters in the model, these indices are expected to reside in the range $-1.8\lesssim \alpha\lesssim 0$ and $-2.2\lesssim \beta \lesssim -4.5$.
Another observational test for any GRB model involving magnetically dominated jets, has to do with the emitted fluxes in the optical and X-ray bands, as compared with observational upper limits. \cite{Beniamini2014} have shown that one zone magnetic models over-produce synchrotron emission in these environments. As is shown in Figs. 4,5 of \cite{Beniamini2014} the typical radius and electrons' Lorentz factor at the photosphere in the model considered here ($r_{ph}\approx 2\times 10^{12}\mbox{cm},\gamma_m\approx 10^3$) are consistent with the upper limits on the optical and X-ray fluxes from observations (this is mainly due to the spectrum being self absorbed below a few keV). Furthermore, due to the gradual energy release in this model, a significant fraction of radiation is produced below the photosphere and re-processed to a thermal form, so the very short cooling times associated with the synchrotron process in these conditions (as compared with the dynamical one) do not imply that synchrotron dominates the peak of the emission, thus overcoming the constraints presented in \citep{Beniamini2013,Beniamini2014} for one zone models.

The basic model presented here consists of relatively few parameters, with rather narrow ranges:
$0.1 \lesssim L_{52} \lesssim 10, 300\lesssim\eta \lesssim 3000,3\times 10^7\mbox{cm}\lesssim \lambda/\epsilon \lesssim 3\times 10^9\mbox{cm},0.1\lesssim \epsilon_e \lesssim 0.5,0.03\lesssim \xi \lesssim 1$ where the latter two are informed by the latest PIC simulations. The model seems to be able to account for the basic prompt GRB observations. Simulations treating the jet dynamics, particle acceleration and radiation self-consistently is required in order to refine the model predictions. This task can, however, be much more involved and is deferred to future works.

We thank Jonathan Granot, Robert Mochkovitch, Frederic Daigne and Tsvi Piran for helpful comments and suggestions.
PB is supported by a Chateaubriand fellowship.

\appendix
\numberwithin{equation}{section}

\section[A] {Approximate estimate of Comptonization effects}
\label{sec:Comptonization}
Inverse Compton of the initially thermal (black body) photon spectrum produced below the photosphere by relativistic electrons accelerated in the reconnection sites, can lead to a modification of the black-body spectrum in a process known as Comptonization.
The importance of Comptonization effects can be gauged by estimating the effective Compton $Y$ parameter that
is felt by the incident photons.
As we are interested in providing an upper limit on the effectiveness of Comptonization, we can safely assume that IC takes place in the Thomson regime and neglect KN effects, that could only reduce the optical depth and energy transfer by IC (and hence the effective Compton parameter).

Since electrons are fast cooling due to synchrotron ($\gamma_c<\gamma_m$), the evolved energy spectrum of the electrons at a given radius is
\begin{equation}
\frac{dn}{d\gamma}\propto \left\{
  \begin{array}{l l}
    \gamma^{-2}, & \quad \gamma_c<\gamma<\gamma_m\\
    \gamma^{-p-1}, & \quad \gamma_m<\gamma\\
  \end{array} \right.
\end{equation} 
The $Y$ parameter is given by:
\begin{equation}
Y(r)=\frac{4}{3}\tau_{IC} \langle \gamma^2 \rangle
\end{equation}
where 
\begin{equation}
\label{eq:gammasquared}
\langle \gamma^2 \rangle=\frac{\int_{\gamma_c}^{\infty} \frac{dn}{d\gamma} \gamma^2 d\gamma}{\int_{\gamma_c}^{\infty} \frac{dn}{d\gamma} d\gamma}\approx \gamma_m \gamma_c \frac{p-1}{p-2}
\end{equation}
and $\tau_{IC}$ is the optical depth for IC scatterings and is given by
\begin{equation}
\label{eq:tauIC}
\tau_{IC}=\int_r^\infty \Gamma (1-\beta)\sigma_{TS} n'_e dr\approx \frac{\xi\sigma_T\dot{M}}{m_pr c \Gamma^2}\bigg(\frac{r}{r_s}\bigg)^{1/3}
\end{equation}
where $n'_e(r)$ is the number of accelerated electrons up to the radius $r$ (in the co-moving frame).
Putting everything together, we find:
\begin{equation}
\label{eq:Y}
Y(r)=0.01 \bigg(\frac{r}{r_{ph}}\bigg)^{2/3}\eta_3^{-4/3}\bigg(\frac{\lambda}{\epsilon}\bigg)_8^{-2/3}\bigg(\frac{\epsilon_e}{0.2}\bigg).
\end{equation}
We find that $Y(r)\ll 1$ for $r < r_{Ph}$.
Comptonization effects are therefore expected to be energetically sub-dominant and cannot significantly change the typical temperature and flux. Numerical simulations are required to gain an accurate estimate of the actual spectrum above the thermal peak, and are deferred to a future work.
The physical situation is very different if one assumes that the dissipated energy smoothly and continuously heats {\it all} the electrons in the flow. In that case the Compton parameter is $Y\sim 1$ for $\tau \sim 1$ and the Compton spectral distortion is pronounced \citep{Giannios2006,Giannios2008}.

It should be noted, that depending on the model parameters, it is possible to formally have $\gamma_m \gamma_c <1$. In this case,
the typical Lorentz factor of the scattering electrons cannot be obtained from Eq. \ref{eq:gammasquared}.
However, we note that this does not change significantly the estimate of $Y(r)$ given by Eq. \ref{eq:Y}.
The reason for this is that since $dn/d\gamma\propto \gamma^{-2}$ for $\gamma<\gamma_m$ and
$Y\propto \int  \gamma^2  dn/d\gamma d\gamma$, we see that different logarithmic bins of $\gamma$ up to $\gamma_m$ all contribute roughly
equally to the overall Compton parameter. Therefore electrons with $\gamma_m$ contribute roughly the same as those, for instance, with $\gamma=\sqrt{\gamma_m \gamma_c}$, and we do not rely on the latter to obtain the result in Eq. \ref{eq:Y}.
Stating it differently, the density of electrons with a Lorentz factor of order $\gamma_m$ is smaller than the total number of accelerated electrons by the ratio of their cooling time to the dynamical time-scale
\begin{equation}
n'_e(\gamma_m)\approx \frac{dn}{d\gamma}\bigg|_{\gamma_m} \gamma_m\approx n'_e \frac{t_{c}(\gamma_m)}{t}=n'_e \frac{\gamma_c}{\gamma_m}.
\end{equation}
Therefore the effective optical depth for IC scatterings with these electrons is reduced by a factor $\gamma_c/\gamma_m$ as compared with Eq. \ref{eq:tauIC}. At the same time, their Lorentz factor squared is simply $\gamma_m^2$ which is a factor $\gamma_m/\gamma_c$ larger than the value in Eq. \ref{eq:gammasquared}. Taking the product of the two, it is clear that $\tau$ remains unchanged and is independent of the actual value of the Lorentz factor for cooled electrons, so long as the electrons are strongly fast cooling.
\end{document}